\documentclass[aps,pra,twocolumn,superscriptaddress]{revtex4-2}
\usepackage[utf8]{inputenc}
\usepackage[T1]{fontenc}

\usepackage{mathbbol}
\usepackage{amsmath}
\usepackage{tikz}
\usepackage{centernot}

\usepackage[utf8]{inputenc}
\usepackage[english]{babel}

\usepackage{amsfonts}
\usepackage{amsthm}
\usepackage{mathrsfs}
\usepackage{mathtools} 

\usepackage{tabularx}
\usepackage{physics,subfigure}

\usepackage{tkz-berge}
\usetikzlibrary{arrows}
\usetikzlibrary{graphs,graphs.standard,fit,shapes.geometric,calc,hobby}

\usepackage{soul}

\begin{document}
\title{Critical metrology of minimally accessible anisotropic spin chains}


\author{Marco Adani}
\affiliation{Center for Cyber-Physical Systems (C2PS), Khalifa University, Abu Dhabi 127788, United Arab Emirates}
\affiliation{Dipartimento di Scienze Fisiche, Informatiche e Matematiche, Universit\`{a} di Modena e Reggio Emilia, I-41125 Modena, Italy}
\author{Simone Cavazzoni}\email{simone.cavazzoni@unimore.it}
\affiliation{Dipartimento di Scienze Fisiche, Informatiche e Matematiche, Universit\`{a} di Modena e Reggio Emilia, I-41125 Modena, Italy}
\author{Berihu Teklu}\email{berihu.gebrehiwot@ku.ac.ae}
\affiliation{College of Computing and Mathematical Sciences, Khalifa University, 127788, Abu Dhabi, United Arab Emirates}
\author{Paolo Bordone}\email{paolo.bordone@unimore.it}
\affiliation{Dipartimento di Scienze Fisiche, Informatiche e Matematiche, Universit\`{a} di Modena e Reggio Emilia, I-41125 Modena, Italy}
\affiliation{Centro S3, CNR-Istituto di Nanoscienze, I-41125 Modena, Italy}
\author{Matteo G. A. Paris}\email{matteo.paris@fisica.unimi.it}
\affiliation{Quantum Technology Lab, Universit\`{a} degli Studi di Milano, I-20133 Milano, Italy}


\date{\today}
\begin{abstract}
We address quantum metrology in critical spin chains with anisotropy and Dzyaloshinskii-Moriya (DM) interaction, and show how local and quasi-local measurements may be exploited to characterize global properties of the systems. In particular, we evaluate the classical (magnetization) and quantum Fisher information of the relevant parameters for the density matrix of a single spin and that of a pair of spins ranging from nearest to sixth-nearest neighbors, to the limiting case of very distant spins. Our results allow us to elucidate the role of the different parameters and to individuate the optimal working regimes for the precise characterization of the system, also clarifying the effects of correlations on the estimation precision.
 \end{abstract}

\maketitle
\section{Introduction}
\label{sec:Introduction}
Quantum phase transitions (QPTs) in many-body systems occur at zero temperature driven by variations of the couplings and/or of external parameters \cite{Sachdev_2000}. The relationship between {both classical and quantum} phase transitions (PTs) and estimation theory is profound and indissoluble. Near (Q)PTs,{(quantum)} states may be discriminated with a high accuracy, because of the enhanced Quantum Fisher Information (QFI) \cite{zanardi2008quantum}. Associated with these collective phenomena, extremely precise detectors can be developed \cite{macieszczak2016dynamical,song2017quantum,garbe2020critical}. In this framework, the Heisenberg XY model with anisotropy and 
Dzyaloshinskii-Moriya (DM) interaction describes a wide range of physical systems, from relatively simple Hamiltonians such as the Ising model to more complex systems that account anisotropy interactions and spin orbit coupling. Moreover, such a model is known for showing multiple phase transitions, making it a good candidate model to test metrological protocols or schemes to detect quantum phase transitions. \cite{jafari2008phase,messio2010schwinger,garate2010interplay,parente2015anomaly,marzolino2017fisher,jin2017phase,yi2019criticality,thakur2020factorization,japaridze2021magnetic,fumani2021quantum,cavazzoni24}. 

In this paper, we demonstrate that local or quasi-local measurements, performed on only one or two spins,  may  be exploited to detect quantum phase discrimination and to precisely characterize the system\cite{pa1,pa2,pa3}. More specifically, we evaluate the classical (magnetization) and quantum Fisher information matrices of the relevant parameters{of the} density matrix of a single spin and that of a pair of spins. {Additionally, we investigate how quantum correlations influence the optimal distance between the two spins, with the goal of maximizing the QFI and thereby optimizing the estimation of the Hamiltonian parameters.}

This model has been computationally investigated \cite{koretsune2018first,cardias2020first,ham2021dzyaloshinskii,mahfouzi2021first,morshed2021tuning,solovyev2023linear}, and found applications in studying materials of technological interest \cite{allwood2005magnetic,imre2006majority,hrabec2020synthetic}, experimental implementations \cite{zhao2021dzyaloshinskii}, and in describing surface and interface phenomena \cite{gusev2020manipulation,akanda2020interfacial,park2020interfacial,legrand2022spatial}. Moreover, DM anisotropy recently found also application in quantum information and technology \cite{shi2017robust,son2019unconventional,ozaydin2020parameter,khlifi2020quantum,houcca2022quantum,motamedifar2023entanglement,zhu2023effect}. Our analysis 
{extends} the application of the model to quantum metrology, and paves the way to realize precise extended sensors requiring only {(quasi-)}local readout schemes.  

The manuscript is structured as follows: In Sec.\ref{sec:DM_interaction}, we introduce the physical model and the theoretical methodologies adopted to investigate the phases for local and quasi-local measurements. In Section \ref{sec:FI_&_QFI}, we introduce the basics of information theory and the quantities monitored to theoretically and practically discriminate the phases of the physical model. Sections \ref{sec:Local_measurements} and \ref{sec:Quasi_local_measurements} present the results of the analysis, distinguishing between measurements of local properties of the system and measurements that involve correlations among the elements of the system. Finally, in Section \ref{sec:Conclusions}, we summarize the main results and highlight differences between the methodologies adopted. {In Appendix \ref{app:appendix_A} and \ref{app:appendix_B} we provide further mathematical details and case studies to support our results.}

\section{Model and methods}
\label{sec:DM_interaction}
The Hamiltonian for an anisotropic XY spin-half chain in the presence of the Dzyaloshinskii-Moriya (DM) interaction can be expressed as:

\begin{align}
	\label{eq:H_syst}
	\mathcal{H} = \sum_{l=1}^{N} & \bigg\{J\big[(1+\gamma)\,\sigma_{l}^{x}\sigma_{l+1}^{x}+(1-\gamma)\,\sigma_{l}^{y}\sigma_{l+1}^{y}+ \nonumber \\
       &D\,(\sigma_{l}^{x}\sigma_{l+1}^{y}-\sigma_{l}^{y}\sigma_{l+1}^{x})\big]-\sigma_{l}^{z}\bigg\}.
\end{align}
where $N$ is the total number of spins and $\sigma_{l}^{x}$,  $\sigma_{l}^{y}$, $\sigma_{l}^{z}$ are the Pauli matrices for the $l$-th spin. $J$ is the coupling constant, and is assumed to be in units of the external magnetic field $B$, with $J=J_{s}/B$, where $J_{s}$ is the actual value of the coupling. The choice to normalize $J$ by $B$ is made because interesting phenomena in the system occur when it is immersed in an external magnetic field. The anisotropy term is represented by $\gamma$ (with $-1 \le \gamma \le 1$),  and $D$ is the Dzyaloshinskii-Moriya (DM) interaction factor. This sets the stage for the rest of the paper, particularly for local and quasi-local approaches to metrology. In both cases, we need the reduced density matrices, which are essential for understanding the system's behavior when only partial accessibility to the system is available. In particular, we are going to consider the thermodynamic limit $N\rightarrow\infty$ and at zero temperature. 
For a local measurement, the single-spin reduced density matrix $\rho_1$ is obtained by tracing out all spins except one from the total density matrix. 
The resulting density operator is actually independent of the specific spin 
and reads 
\begin{equation}
	\label{eq:reduced DM 1}
	\rho_1 = \frac12 \left ({\mathbb I}+ \langle \sigma^{z}\rangle\, \sigma^z\right)
\end{equation}
where $\langle\sigma^{z}\rangle =\hbox{Tr}[\rho_1\, \sigma^z]$ denotes the mean magnetization per spin 
of the system, i.e.
\begin{equation}
   \label{eq:magnetization}
     \langle \sigma^z \rangle = - \frac{1}{\pi} \int_0^{\pi} d\phi \ \frac{\left[J(\cos \phi - 2D \sin \phi)-1\right]}{\Delta},
\end{equation}
where 
\begin{equation}
    \Delta = \sqrt{\left[ J(\cos \phi - 2D \sin \phi)-1 \right]^2 + J^2 \gamma^2 \sin^2 \phi}.
\end{equation}
Indeed, the reduced density matrix of a single spin only depends on the local properties of the system: the diagonal elements depends only on the local magnetization, and the reduced density is the same independently on the exact location of the spin, as expected from the translational invariance of the Hamiltonian in Eq.\eqref{eq:H_syst}.

If we move to two spins, the reduced density matrix includes not only local properties but also the correlations between them. Upon tracing over all the spins except those corresponding to the $j$ and $k$ sites we have
\begin{equation}
    \label{eq:reduced 2}
     \rho_2=\hbox{Tr}_{\overline{jk}}(\rho).
\end{equation}
Due to the symmetries inherent in the physical model under analysis, the two-site reduced density matrix has the following 
$X$-structure \cite{radhakrishnan2017quantum} 
\begin{equation}
	\label{eq:reduced DM 2}
	\rho_2 \equiv \rho_2 (r) =
\begin{pmatrix}
 	a_+ & 0 & 0 & b_- \\
	0 & c & b_+ & 0 \\
	0 & b_+ & c & 0 \\
	b_- & 0 & 0 & a_-  
\end{pmatrix},
\end{equation}
where the matrix elements $a_{\pm}$, $b_{\pm}$ and $c$ are given by 
\begin{equation}
    \begin{split}
    a_{\pm}=&\frac{1}{4}\big[ 1 \pm 2 \langle \sigma^z \rangle +\langle \sigma^z_j  \sigma_{j+r}^z \rangle \big], \\
    b_{\pm}=&\frac{1}{4}\big[  \langle \sigma_j^x \sigma_{j+r}^x \rangle \pm \langle \sigma_j^y  \sigma_{j+r}^y \rangle  \big], \\
    c=&\frac{1}{4} \big[ 1 - \langle \sigma_j^z \sigma_{j+r}^z \rangle \big].
    \end{split} 
\end{equation}
The quantity $\langle\sigma^{z}\rangle$ is given in Eq.\eqref{eq:magnetization} whereas 
\[
\langle \sigma_j^\alpha \sigma_{j+r}^\alpha \rangle\equiv S^\alpha_r\, \quad \alpha=x,y,x \]
denote the correlations between the components of the two spins. Due to the translational invariance of the Hamiltonian in Eq.\eqref{eq:H_syst}, these correlation functions do depend only on the distance $r$ between the two spins. The spin-spin correlation functions $S^x_r$ and $S^y_r$ can be computed from the determinant of Toeplitz matrices \cite{liu2011quantum,liu2017quantum} as
\begin{equation}
   S^x_r = 
    \begin{vmatrix}
    	G_{-1} & 	G_{-2} & \dots & 	G_{-r} \\
    	G_{0} & 	G_{-1} & \dots & 	G_{-r+1} \\
    	\vdots &   \vdots &   \ddots &  \vdots \\
    	G_{r-2} & 	G_{r-3} & \dots & 	G_{-1} 
    \end{vmatrix} \ ,
\end{equation}
\begin{equation}
    S^y_r = 
    \begin{vmatrix}
    	G_{1} & 	G_{0} & \dots & 	G_{-r+2} \\
    	G_{2} & 	G_{1} & \dots & 	G_{-r+3} \\
    	\vdots &   \vdots &   \ddots &  \vdots \\
    	G_{r} & 	G_{r-1} & \dots & 	G_{1} 
    \end{vmatrix} \ ,
\end{equation}
in which $r$ is the distance between the two spins. For the z-direction we have 
\begin{equation}
  S^z_r = {\langle {\sigma}_{i}^z \rangle}^2 - G_r G_{-r} \ ,
\end{equation}
where
\begin{align}
  \label{eq:G_k}
    G_k =& - \frac{1}{\pi} \int_0^{\pi} d\phi \ \frac{2 \cos (\phi k)}{\Delta} \left[J(\cos \phi - 2D \sin \phi)-1\right] \nonumber \\
    & + \frac{\gamma}{\pi} \int_0^{\pi} d\phi \ \frac{2 J \sin (\phi k)}{\Delta} \sin{\phi}.
\end{align}

\section{Quantum and Classical Information Theory}
\label{sec:FI_&_QFI} 
As the aim of this work is to focus on metrology under the assumption of partial system accessibility, it is crucial to introduce key quantities that serve to distinguish between various states of the model. The study is based on the analysis of the quantum and classical Fisher information, which quantify the ultimate bounds to precision in the estimation of system parameters \cite{giorda1,giorda2,teklu3,PhysRevA.92.010302}. The QFI is intrinsically related to the geometry of the manifold of quantum states, i.e. the Bures distance 
\cite{sommers2003bures,paris2009quantum} as
\begin{equation}
    \label{eq:bures_metric}
    d^{2}_{B}(\rho_{\lambda},\rho_{\lambda+d\lambda}) = \frac{1}{4} H(\lambda) d\lambda^{2},
\end{equation}
where $H(\lambda)$ is referred to as the quantum Fisher information  (QFI)\cite{gu2010fidelity,damski2016fidelity,vsafranek2017discontinuities} associated to a parameter $\lambda$ of the Hamiltonian. The QFI is itself related to the fidelity \cite{gu2010fidelity,damski2016fidelity,vsafranek2017discontinuities}
and may be evaluated as
\begin{equation}
    \label{eq:general_qfi}
    H(\lambda) = \hbox{Tr}[ \rho_{\lambda}\mathcal{L}_{\lambda}^{2} ],
\end{equation}
where $\mathcal{L}$, the symmetric logarithmic derivative, is related to the variation with respect to a parameter $\lambda$ of the density matrix $\rho$ as
\begin{equation}
    \label{eq:sld}
    \partial_{\lambda} \rho_{\lambda} = \frac{1}{2} \{ \mathcal{L}_{\lambda}, \rho_{\lambda} \},
\end{equation}
where $\{ \ , \ \}$ indicates the anti-commutator. 

The QFI sets a  bound on the variance of any (unbiased) estimator used to infer the value of the parameter of interest from data, as 
\begin{equation}
    \label{eq:qvar_var}
	V(\lambda) \geq \frac{1}{MH(\lambda)},
\end{equation}	
where $M$ is the number of measurements. Since quantum phase transitions are described as an abrupt change in the ground state of a many-body system due to the variation of a physical parameter, there is a deep relation among phase transitions and the QFI \cite{invernizzi2008optimal,zanardi2008quantum,sun2010fisher,carollo2020geometry,chu2021dynamic,mihailescu2023}, and phase transitions are connected to divergences in the QFI. In particular, we expect critical spin chains to provide enhanced precision for those values of the parameters corresponding to quantum phase transitions, where $H(\lambda)$ diverges.

The single-spin density matrix in Eq. (\ref{eq:reduced DM 1}) is diagonal in the basis of $\sigma^z$, i.e. $\rho_1= p |0\rangle\langle 0|+ (1-p) |1\rangle\langle 1|$ with $p = \frac12 (1+ \langle \sigma^z \rangle)$ and thus the QFI may be easily evaluated as 
\begin{equation}
    \label{qfisingle}
	H(\lambda) = \frac{(\partial_\lambda p)^2}{(1-p) p} = 
	\frac{(\partial_\lambda \langle \sigma^z \rangle)^2}{1-\langle \sigma^z \rangle^2}
\end{equation}

When we have more than one parameter, Eq. (\ref{eq:bures_metric}) 
generalizes to 
\begin{equation}
    \label{eq:bures_metricM}
    d^{2}_{B}(\rho_{\boldsymbol{\lambda}},\rho_{\boldsymbol{\lambda}+d\boldsymbol{\lambda}}) = \frac{1}{4} H_{\mu\nu}(\boldsymbol{\lambda}) d\mu d\nu,
\end{equation}
where $H_{\mu\nu}(\boldsymbol{\lambda})$ are the elements of the so-called quantum Fisher information matrix (QFIM), defined as 
\begin{equation}
    \label{eq:general_qfim}
    H_{\mu\nu}(\boldsymbol{\lambda}) = \frac12\,\hbox{Tr}[\rho_{\boldsymbol{\lambda}}
    \left(\mathcal{L}_{\mu}
    \mathcal{L}_{\nu}+
    \mathcal{L}_{\nu}
    \mathcal{L}_{\mu}
    \right)
    ],
\end{equation}
where $\mathcal{L}_{\mu}$ and $\mathcal{L}_{\nu}$ are the symmetric logarithmic derivatives associated to the components $\lambda_{\mu}$ and $\lambda_{\nu}$ of the vector $\boldsymbol{\lambda}$ respectively. 
For the spin chain Hamiltonian (Eq.\eqref{eq:H_syst}) the parameters are the coupling constant $J$, the anisotropy parameter $\gamma$ and the DM interaction factor $D$, so in this case $\boldsymbol{\lambda}=\{J,\gamma,D\}$. 

The inverse of the QFI matrix provides a lower bound on the covariance matrix 
$\boldsymbol{V}$ of the set of estimators, i.e., ${\boldsymbol{V}}_{\mu\nu}=\langle \lambda_{\mu} \lambda_{\nu}\rangle-\langle \lambda_{\mu} \rangle \langle \lambda_{\nu}\rangle$, which reads 
\begin{equation}
    \label{eq:Cov_matr_bound}
    \boldsymbol{V} \boldsymbol{(\lambda)} \geq \frac{1}{M} {\boldsymbol{H}}^{-1}\boldsymbol{(\lambda)},.
\end{equation}
By introducing a positive, real matrix of dimension $n \cross n$, i.e. the weight matrix $W$, a scalar bound may be obtained. In the following we consider simple choice $W=\mathbb{1}$, {obtaining a relation in the form}
\begin{equation}
    \label{eq:Var_lower_bound}
     \hbox{Tr} \left[ {\boldsymbol{WV}} \right]  = \sum_{\mu} \hbox{Var}(\lambda_{\mu})  \geq \frac{1}{M} \hbox{Tr} \left[ {\boldsymbol{H}}^{-1}\boldsymbol{(\lambda)} \right] \ ,
\end{equation}
i.e. a lower bound on the sum of the variances associated to the parameters contained in the vector $\boldsymbol{\lambda}$. For the single-spin density matrix in Eq. (\ref{eq:reduced DM 1}) the elements of the QFIM are given by 
\begin{equation}
H_{\mu\nu} (\boldsymbol{\lambda}) = \frac{\partial_\mu p\, \partial_\nu p}{(1-p)p} = \frac{\partial_\mu \langle \sigma^z \rangle \, \partial_\nu \langle \sigma^z \rangle}{1-\langle \sigma^z \rangle^2}\,.
\end{equation} 

For the two-spin reduced density matrix of Eq. (\ref{eq:reduced DM 2}) the QFI read as follows \cite{maroufi2021analytical} 
\begin{align}
    \label{eq:H_rho1}
    H_r(\lambda) = & \frac{1}{a_0} \left[ \frac{(\sum_{jk}\eta_{jk} a_j \partial_\lambda a_k)^2}{\sum_{jk}\eta_{jk} a_j a_k} 
    - \sum_{jk} \eta_{jk} \partial_\lambda a_j\,  \partial_\lambda a_k \right] \nonumber \\
+ & \frac{1}{b_0} \left[ \frac{(\sum_{jk}\eta_{jk} b_j \partial_\lambda b_k)^2}{\sum_{jk}\eta_{jk} b_j b_k} 
    - \sum_{jk} \eta_{jk} \partial_\lambda b_j\,  \partial_\lambda b_k \right] \nonumber \\    
    + & \frac{(\partial_\lambda a_0)^2}{a_0} + \frac{(\partial_\lambda b_0)^2}{b_0}\ .
\end{align}
where
\begin{align}
a_0 &= \frac{1}{2}\left( 1 + S_r^z\right),\, a_1  =\frac{1}{2}\left( S_r^x -S_r^y\right),\, a_2 = 0,\, a_3 =  \langle \sigma^z \rangle\,,\notag \\
b_0 &= \frac{1}{2}\left( 1 - S_r^z\right),\, b_1  =\frac{1}{2}\left( S_r^x + S_r^y\right),\, b_2 = 0,\, b_3 =  0\,.
\end{align}
and $\eta = \hbox{Diag}\{1,-1,-1,-1\}$.

The QFI provides an upper bound to the Fisher information (FI) of any possible measurement that may be performed on the system in order to infer the value of the parameter. The FI itself is defined as 
\begin{equation}
    \label{eq:general_FI}
	F(\lambda) = \sum_{x} p(x \vert \lambda) \big[ \partial_{\lambda} \log p(x \vert \lambda) \big]^{2} \,.
\end{equation}
and since the density matrix $\rho_1$ is diagonal in the $z$-basis, 
we have that for local magnetization measurement $F(\lambda)=H(\lambda)$.
For a two spin magnetization measurement the FI is given by
\begin{equation}
    \label{eq:FI_system_2}
	F(\lambda)= \frac{1}{a_+} \left(\frac{\partial a_+}{\partial \lambda} \right)^2 + \frac{2}{c} \left(\frac{\partial c}{\partial \lambda} \right)^2 + \frac{1}{a_-} \left(\frac{\partial a_-}{\partial \lambda} \right)^2 \,,
\end{equation}
and since $\rho_2$ is non-diagonal in the $z$-based, the Fisher information 
is lower than its quantum counterpart, i.e., 
\begin{equation}
    \label{eq:FI_QFI}
 F(\lambda)\leq   H(\lambda)\,.
\end{equation}
Another relevant feature associated to the Fisher matrix is its determinant. It is a measure of the \textit{degree of sloppiness of the system} that quantify how strong is the dependence of the system on a combination of the components of $\boldsymbol{\lambda}$ rather then on its components separately. When the Fisher matrix is singular (i.e. $det[\boldsymbol{H(\lambda)}]$=0), the statistical model is referred to as \textit{sloppy}, and this means that the true parameters that describe it are combinations of the original parameters $\{{\lambda}_{1},{\lambda}_{2},...,{\lambda}_{n}\}$. For this reason the closer is $det[\boldsymbol{H(\lambda)}]$ to zero the higher is the degree of sloppiness of the system. 

Finally, a fundamental tool in multi-parameter estimation is the so called Uhlmann matrix, whose elements are defined as 
\begin{equation}
    \label{eq:general_Uhlmann_matrix}
    {\boldsymbol{U}}_{\mu\nu}\boldsymbol{(\lambda)} = \hbox{Tr}\left[\rho_{\boldsymbol{\lambda}}\frac{\mathcal{L}_{\mu}\mathcal{L}_{\nu}-\mathcal{L}_{\nu}\mathcal{L}_{\mu}}{2}\right].
\end{equation}
A vanishing Uhlmann matrix means that the parameters may be jointly estimated with the same precision achievable from their separate estimations. On the other hand, if ${\boldsymbol{U}}_{\mu\nu}\boldsymbol{(\lambda)}\neq0$ there is an intrinsic additional noise of quantum origin in the joint estimation of $\lambda_{\mu}$ and $\lambda_{\nu}$, due to the non commutativity of the corresponding SLDs. 

The integrals in Eq.\eqref{eq:magnetization} and Eq.\eqref{eq:G_k} cannot be evaluated analytically and therefore we study the (Q)FI {and all the quantities related to them} numerically. Moreover, given the structure of the Hamiltonian  in Eq.\eqref{eq:H_syst}, we study the phenomenology of the (Q)FI for $J \neq 0$.

\section{Local measurements}
\label{sec:Local_measurements}
We start our analysis from the information that can be extracted by accessing a single spin. In this case, the best measurement is the magnetization along the direction of the external field since, as mentioned above, its FI equals the QFI. 
\begin{figure}[h!]
    \centering
    \includegraphics[width=0.9\columnwidth]{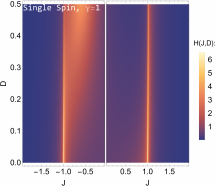}
    \caption{(Quantum) Fisher Information of $J$ for a single spin reduced density matrix of an anisotropic Heisenberg XY spin chain, with Dzyaloshinskii-Moriya interaction. The anisotropy parameter value is fixed to $\gamma=1$ (Ising model). Here we can see the dependence on the Hamiltonian parameters $J$ and $D$, for $J\neq0$. For $J=\pm1$, $H(J)$ diverges $\forall D$. This allows to distinguish the two phases of the system from a local measurement.}
    \label{fig:Single_spin_ising_HM}
\end{figure}

In Fig.\ref{fig:Single_spin_ising_HM} we show the QFIM element $H_{JJ}(J,D)$ associated to the coupling constant of the spin chain as a heat-map, as a function of $J$ and $D$ for a fixed value of $\gamma=1$ (Ising model). We notice a divergence for $J=\pm 1$, as it happens for the QFI of a collective measurement \cite{invernizzi2008optimal}. The parameter $D$ affects the behavior of $H_{JJ}$ mainly for $J<0$. As $D$ increases, the range of $J$ in which $H_{JJ}$ is large increases too, whereas the value at the peak in $J=-1$ decreases. Results for $\gamma \neq 1$ are similar.
These results show how that due to the Hamiltonian properties, even a local measurement detects the phase transitions of the spin chain and therefore may be exploited in a metrological protocol. The precision should be however be compared to that achievable by quasi local measurements, where correlations usually enhance the information that may extracted. This will be the subject of the next Section.

\section{Quasi local measurements and the role of correlations}
\label{sec:Quasi_local_measurements}
Let us now consider estimation protocols based on measuring two spins of the chain. In this case, besides local magnetization, the results of the measurement are also influenced  by the correlations between the spins. To begin, it is useful to start from the limit of infinitely distant spins. This because for 
non-interacting spins the (Q)FI is just twice the single-spin (Q)FI, and this case may be used  as a reference to understand whether {correlations} are beneficial or detrimental for quantum metrology. Since the Hamiltonian involves interactions between nearest neighbors, we may expect the correlation to vanish by increasing the distance between the two measured spins. Indeed, we found numerically that $|G_{r}|<1$ $\forall r$ for the entire range values of $J$, $\gamma$ and $D$ used in this work. Furthermore, we found that the functions $G_{r}$ goes to zero as $1/r$.
\begin{equation}
      \label{eq:G_r inf dist}
      G_{r}  \simeq -G_{-r}  \propto \frac{1}{r} \ \ \hbox{for} \ \ r \rightarrow \infty \,,
\end{equation}
and, in turn, 
\begin{align}
   \label{eq:Correlations_r_inf}
      \langle \sigma_i^x \sigma_{i+r}^x \rangle & \propto \frac{1}{r}  \qquad
      \langle \sigma_i^y \sigma_{i+r}^y \rangle  \propto \frac{1}{r} \\
      \langle \sigma_i^z \sigma_{i+r}^z \rangle & \simeq {\langle {\sigma}_{i}^z \rangle}^2 \,.
\end{align}
Correspondingly, the two-spin reduced density matrix \eqref{eq:reduced DM 2} assumes the diagonal form 
\begin{equation}
	\label{eq:reduced DM inf 1}
	\rho_2(\infty) = \hbox{Diag}[a_+,c,c,a_-]\,.
\end{equation}
This structure implies that the classical FI of magnetization equals the QFI 
and is given by 
\begin{equation}
    \label{FI r inf 1}
	H(\lambda) = F(\lambda) = 2\, \frac{{\partial_\lambda \langle \sigma^z \rangle}^2}{{1-\langle \sigma^z \rangle}^2}\,,
\end{equation}  
i.e., in this case, measuring the magnetization allows us to extract the entire information present in the system, which is twice the value of the single spin case. This is true $\forall \lambda \in \{J,\gamma, D\}$. In turn, this results follows directly from the symmetry of the model since, due to the translation invariance of the Hamiltonian in Eq.\eqref{eq:H_syst} every spin of the model contributes equally to the (Q)FI in the absence of correlations.

We now move to the study of the FI and QFI as a function on the distance between the measured spins, looking for optimal configurations in the different realizations of the model.
Let us start from the Ising model without DM interaction (i.e. the anisotropy parameter is set to $\gamma=1$ and the DM interaction factor to $D=0$).
\begin{figure}[h!t]
        \centering
        \includegraphics[width=0.96\columnwidth]{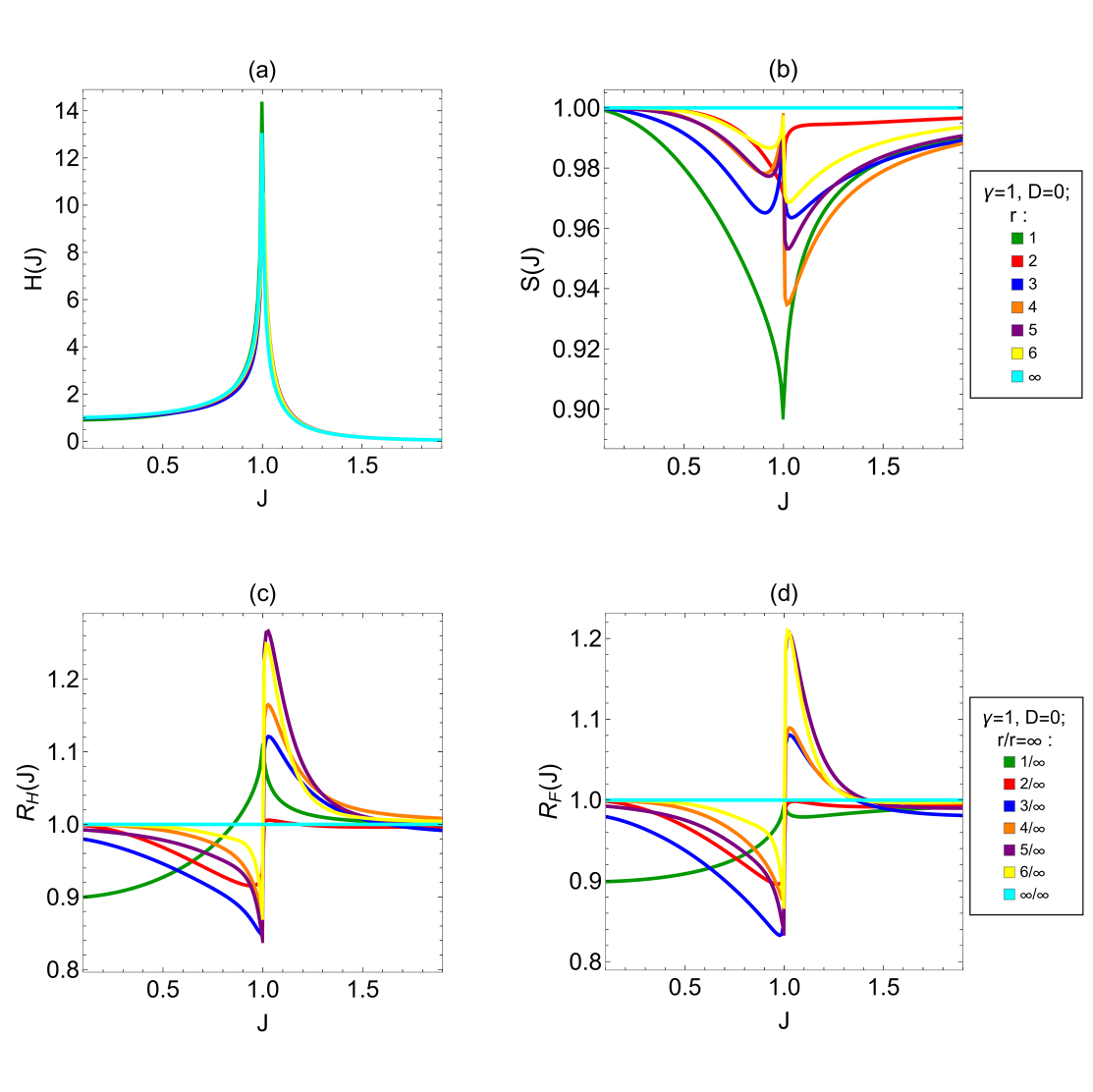}
        \caption{--(a) QFI of coupling constant $J$ for different values of $r$. (b) Saturation associated to $J$. $S(J)$ is always above 0.9. In the limit of infinitely distant spins $H(J)=F(J)$. (c)-(d) Ratios between the (Q)FI for different distances and the limiting QFI of infinitely distant neighbors $R_{(H)F}(J)$. For $H(J)$ and $F(J)$ the optimal distance depends differently on $J$. --The anisotropy parameter is $\gamma=1$ and the DM factor is $D=0$. Different curves and colors are associated to different values of the distance between the two spins measured $r$.}
        \label{fig:Ising_D0_quasilocal}
\end{figure}
In Fig.\ref{fig:Ising_D0_quasilocal}, we show the QFI $H(J)$ as a function of $J$ for different distances between the pair of spins. We show $H(J)$ only for $J$ positive, because when $D=0$ the QFI is even in $J$, $H(J)=H(-J)$. The 
behavior of $H(J)$ is qualitatively the same regardless the distance. In particular all curves show a divergence for $J=\pm1$. This implies that $H(J)$ is strongly affected by the phase transition from the ferromagnetic to the paramagnetic phase of the spin chain and the system is an excellent sensor in this region. To understand if the phase transition is detectable in practice, we also studied the FI $F(J)$ of magnetization measurement. The general behavior of $F(J)$ and $H(J)$ is very close, with the same symmetry, $F(J)=F(-J)$ and the same divergences at $J=\pm 1$, though they differ quantitatively. For this reason, we show the ratio between these two quantities, termed saturation and defined as 
\begin{equation}
    \label{eq:saturation}
    S(J)=\frac{F(J)}{H(J)}
\end{equation}
In the top right panel of Fig.\ref{fig:Ising_D0_quasilocal}, we show $S(J)$ for different $r$. As mentioned above, $S(J)=1$ for $r\rightarrow \infty$, and it remains very high ($S(J) \gtrsim 0.9$) for the entire range of distances we explored ($1\leq r \leq 6$). We conclude that is feasible to extract a large part of information from the Ising spin chain through a magnetization measurement. To understand if there is an optimal distance that maximizes the QFI and the FI we look at the ratios between the (Q)FI of the different neighbors and (Q)FI in the limit of infinitely distant spins, i.e. the quantities
\begin{equation}
    \label{eq:ratios}
    R_{H}(J)=\frac{H(J;r)}{H(J;\infty)} \; ; \; R_{F}(J)=\frac{F(J;r)}{F(J;\infty)}\,,
\end{equation}
which is shown in the lower panels of Fig.\ref{fig:Ising_D0_quasilocal}. 
From the lower left panel (i.e. QFI ratios), we see that the optimal distance between the measured spins depends on the value of $J$ itself, and in turn on the value of the external magnetic field.  In the ferromagnetic phase (i.e. $J<1$) and for $0<J \lesssim 0.9$, the optimal choice is to measure very distant spins ($r \rightarrow \infty$), whereas in the region from $J\approx0.9$ to $J=1$ we have the opposite, i.e. the optimal choice is  to measure two nearest neighboring spins ($r=1$). In the paramagnetic phase, from $J=1$ to a value close to $J\approx1.25$ the optimal distance is $r=5$, while for higher values of $J$ the optimal distance is $r=4$.  Concerning the FI, we see from the lower right panel of Fig.\ref{fig:Ising_D0_quasilocal} that also in this case 
the optimal distance depends on $J$. 

In the ferromagnetic phase, $J<1$, the optimal distance is $r\rightarrow\infty$. In the paramagnetic phase, in the region close to $J=1$ the optimal distance is $r=6$, then up to a value close to $J=1.4$ it becomes $r=5$ and after this value it goes again to $r\rightarrow\infty$. In this phase, close to $J=1$, we can notice an even/odd effect such that the curves for $r=3$ and $r=4$ and the curves for $r=5$ and $r=6$ are almost paired. A similar effect is present also in the ratios for the QFI, but less marked. Comparing the lower panels of Fig.\ref{fig:Ising_D0_quasilocal}, another emerging feature is that the optimal distances to optimize FI and QFI are in general different for a given value of $J$. 

\begin{figure}[ht!]
    \centering
    \includegraphics[width=0.96\columnwidth]{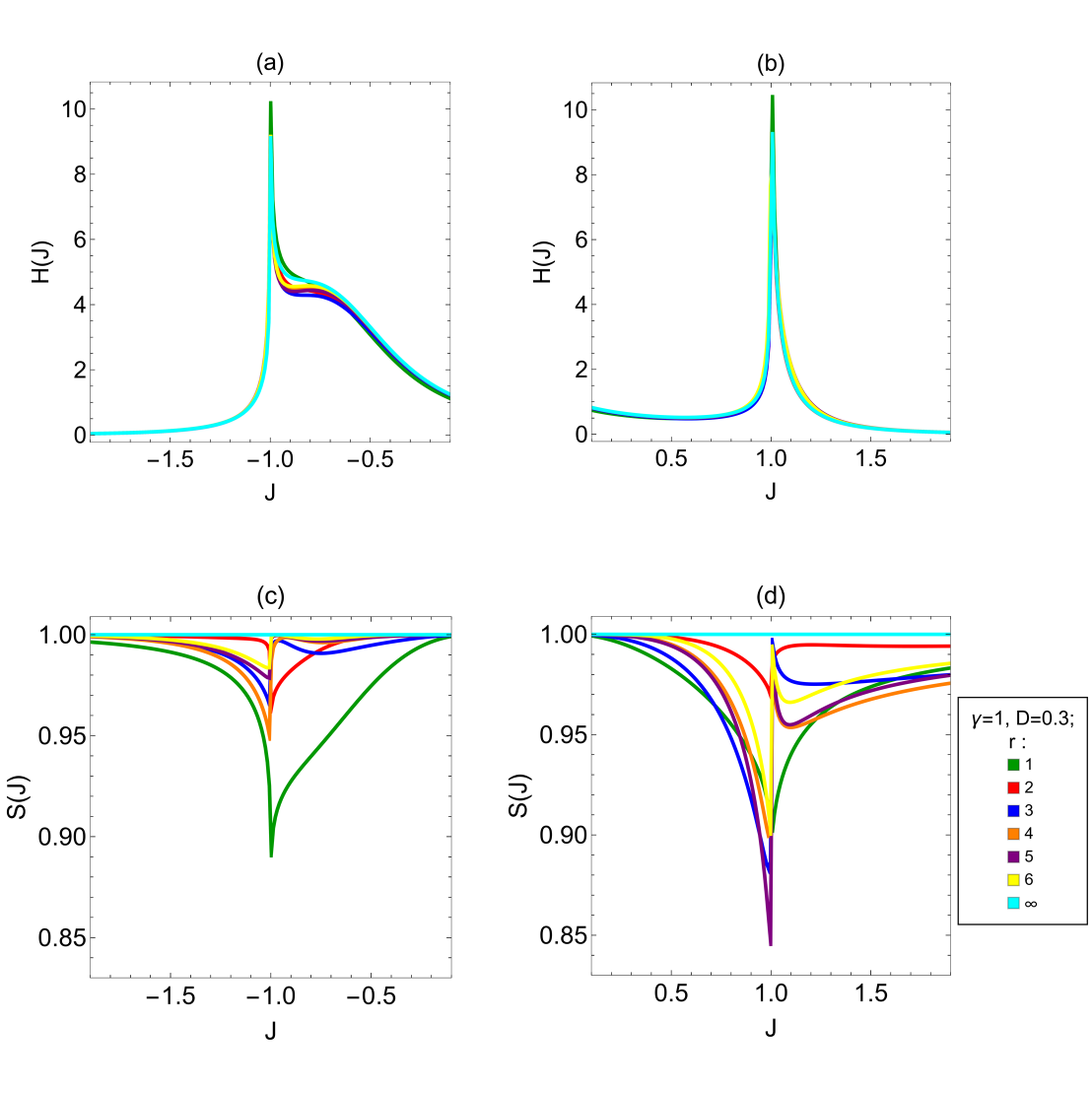}
    \caption{--(a)-(b) QFI of the coupling constant $J$. (c)-(d) Saturation $S(J)$ associated to the coupling constant $J$. In the limit of infinitely distant spins $H(J)=F(J)$. -- All the plots in figure are as $J$ varies. The anisotropy parameter is $\gamma=1$ and the DM factor is $D=0.3$. The different curves are associated to the different values of the distance between the two spins measured $r$.}
    \label{fig:Ising_D03_quasilocal_H_S}
\end{figure}

After the analysis of the Ising model, we now study the impact of the DM interaction. For the sake of concreteness, we focus on the specific case 
of $\gamma=1$ and $D=0.3$. Notice also that due to the structure of the Hamiltonian, what matters is the relative sign between $J$ and $D$ and the phenomenology observed for positive (negative) $J$ and $D>0$ is the same observed for negative (positive) $J$ and $D<0$. In the top panels of Fig.\ref{fig:Ising_D03_quasilocal_H_S}, we show the QFI $H(J)$ as a function of $J$ for different spin distances $r\in\{1,2,3,4,5,6,\infty\}$.
The first thing to notice is that the QFI is no longer even in $J$, whereas 
the general behavior of the curves is nearly independent on $r$, as it happens for $D=0$. The two divergences at $J= \pm 1$ are still present, but in this case a bump appears on the right of the divergence at $J=-1$. These features appear 
also for the FI and in turn, independently on $D$, the FI and the QFI are very close, as it is apparent from the the plot of the saturation $S(J)$, in the 
lower panels of Fig.\ref{fig:Ising_D03_quasilocal_H_S}, which is always above $S(J)\approx 0.84$. 
\begin{figure}[h!]
    \centering
    \includegraphics[width=0.96\columnwidth]{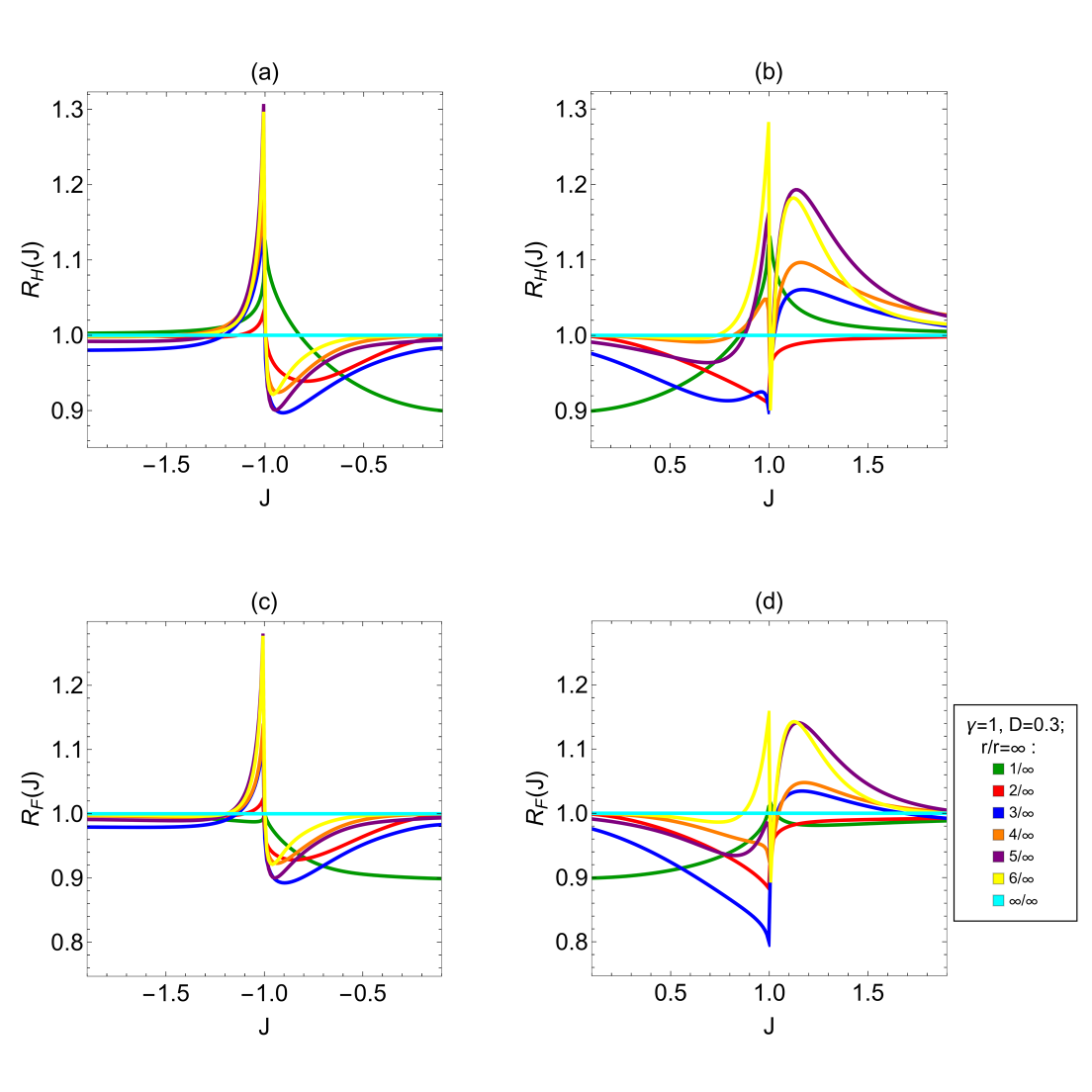}
    \caption{--(a)-(b) Ratios between the QFI for two spins at distance $r$ and QFI in the limit of infinitely distant neighbors, $R_{H}(J)$ as a function of $J$ and for different $r$. As it was without DM interaction, also in this case 
the optimal distance depends on $J$. (c)-(d) Ratios between the FI of the different neighbors and FI in the limit of infinitely distant neighbors, $R_{F}(J)$ as a function of $J$ and for different $r$. -The anisotropy parameter is $\gamma=1$ and the DM factor is $D=0.3$.}
    \label{fig:Ising_D03_quasilocal_RH_RF}
\end{figure}    

In order to find for the optimal spin distance in the presence of DM interaction, we look at the ratios between the QFI (FI) for different spin distances and the that associated to infinite distance {(i.e. $R_{H(F)}(J)$)}, see the top panels of Fig.\ref{fig:Ising_D03_quasilocal_RH_RF}. Also in presence of DM interaction the optimal distance depends on $J$. In the region around the value $J\approx-2$, the optimal distance is $r=1$, then for a very short interval $r=4$ and after up to $J=-1$ it is $r=5$. In the ferromagnetic phase, from $J=-1$ to a value close to $J\approx-0.85$ the optimal distance is $r=1$ and then it is $r=\infty$ up to another value of $J$ close to $J=0.75$. After that the optimal distance is $r=6$ until $J=1$. For $J>1$ the optimal distance is $r=1$ up to a value of $J$ close to $J=1.05$ then it is $r=5$. Comparing these ratios to those for Ising model (Fig.\ref{fig:Ising_D0_quasilocal}), we can see two interesting features arising from the presence of the DM interaction. The first one is that the peaks associated to $r=3$, $r=4$, $r=5$ and $r=6$ in the paramagnetic phase, close to $J=1$, are higher respect to the case without DM interaction for $J<0$ and lower than the case without DM interaction for $J>0$. This is a general trend observed for all the values of $\gamma$ and $D$ studied (see also Appendix \ref{app:appendix_A}). In particular, the higher is the value of $D$, the higher is the increase respect to the case $D=0$. The other interesting feature is that for $J<0$ the curves associated to $r=4$, $r=5$ and $r=6$ are higher than the curve for $r=\infty$. 

Concerning the ratios of FI, the lower panels of Fig.\ref{fig:Ising_D03_quasilocal_RH_RF} show that starting from $J\approx-2$ the optimal distance is $r=\infty$ up to a value of $J$ close to $J\approx-1.2$ then until $J=-1$ it is $r=5$. In the ferromagnetic phase, the optimal distance is $r=\infty$ up to a value of $J$ close to $J=0.85$ after that it is $r=6$ until the phase transition. In the paramagnetic phase, very close to $J=1$ the optimal distance is $r=1$, then it almost immediately changes to $r=6$, whereas approaching  $J=1.1$ it becomes $r=5$. Comparing these ratios with those of the the Ising model , we see the same feature for the peaks in the paramagnetic phase that is observed for the QFI. Also for the ratios of the FI with $D=0.3$ in paramagnetic phase there is a loss in the peaks close to $J=1$ for the curves associated to $r=3$, $r=4$, $r=5$ and $r=6$, respect the analogue curves for $D=0$. Instead there is a gain for the peaks close to $J=-1$. Another interesting feature is that, as for the QFI, in the FI associated to the ferromagnetic phase when $J>0$ the optimal distance close to the phase transition is $r=6$, not $r=\infty$ as it is for the Ising model without DM interaction. 

From the comparison between the results with and without DM interaction, few general conclusions may be drawn. The first is that the optimal distance for the FI and the QFI are, in general, different. This could be expected from the fact that, except for the limit of infinitely distant spins, the QFI depends on all the correlation functions $\langle \sigma_i^x \sigma_{i+r}^x \rangle$, $\langle \sigma_i^y \sigma_{i+r}^y \rangle$ and $\langle \sigma_i^z \sigma_{i+r}^z \rangle$, whereas the FI depends only on $\langle \sigma_i^z \sigma_{i+r}^z \rangle$. In addition, we have that the optimal distance for both the FI and the QFI depends on all the Hamiltonian parameters, $J$, $\gamma$ and $D$. 
\section{Multi-parameter estimation for quasi local measurements}
\label{sec:Multi-Parameter_estimation_for_quasi_local_measurements}
After having analyzed the bounds to precision for the coupling constant $J$, we now move to study the impact of the distance $r$ between the measured  spins on the precision of all Hamiltonian parameters. First of all, we notice that being the density matrix of the system a real X-state the Uhlmann matrix vanishes (see Appendix \ref{app:appendix_B}). Actually, the symmetric logarithmic derivatives for the different parameters do not commute, but they do {\em weakly}, i.e. the commutators have vanishing expectation value, such that the system is \textit{asymptotically classical}\cite{ALBARELLI2020126311}. As a consequence, 
it is possible to perform the joint estimation of the Hamiltonian parameters $J$, $\gamma$ and $D$ without any additional intrinsic noise of quantum origin. 

\begin{figure}[h!]
        \centering
        \includegraphics[width=0.96\columnwidth]{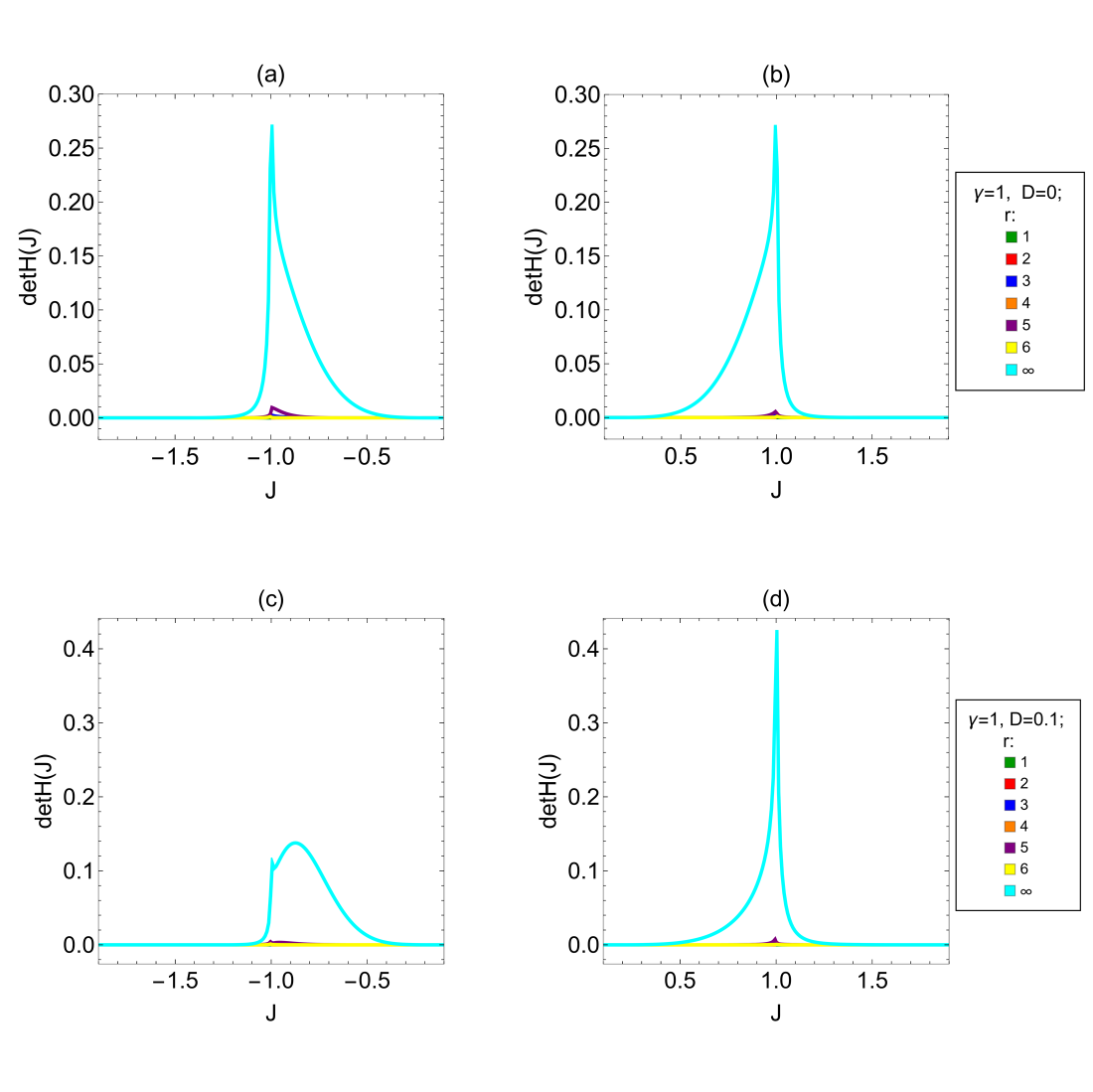}
        \caption{Determinant of the quantum Fisher matrix $\hbox{det}H$ as a 
        function of $J$. --(a)-(b) Ising model without DM interaction. (c)-(d) Ising model with DM interaction and $D=0.1$. -- Different curves and colors are associated to different values of the distance $r$ between the two measured spins; $r=\infty$ correspond to the limit of infinitely distant spins.}
        \label{fig:Ising_determinants}
\end{figure}

Let us now analyze the degree of sloppiness of the model, i.e. whether the state of the system depends on $J$,$\gamma$ and $D$ separately or only on a 
combination of them. This may investigated by looking at the determinant of the QFI matrix, see Fig.\ref{fig:Ising_determinants}.
    
It is immediately clear that, except close to the phase transitions $(J=\pm1)$, for both $D=0$ and $D=0.1$ the determinant is negligible when $1\leq r \leq 6$. This means that when the system is not close to the phase transitions the degree of sloppiness is so high that we can consider it sloppy for any practical purposes. For these values of $r$ even close to $J=\pm1$ the degree sloppiness is still very high. For $r=\infty$, there are wide intervals of $J$ for which the determinant is significantly different from zero. In these regions the degree of sloppiness is much lower then for the other neighbors, therefore $r=\infty$ is the most suitable choice for the practical joint estimation of the Hamiltonian parameters. This behavior holds regardless the value of $D$, which has a weak effect on the specific position and width of the intervals, but not on their presence. This suggests that the correlations among the measured spins influence the degree of sloppiness of the system, and this influence can be so strong to lead to decrease the number of effective parameters of the system. For very distant spins the reduced density matrix is diagonal and the QFIM is not affected by correlations (see Eq.\eqref{eq:reduced DM 1}), and the intervals of $J$ for 
which the system is not sloppy are significantly wider. Nevertheless there are still regions in which the determinant is nearly zero even, and we conclude that 
the degree of sloppiness does not depend only on the {correlations} among the spins. 

\begin{figure}[h!]
        \centering
        \includegraphics[width=0.96\columnwidth]{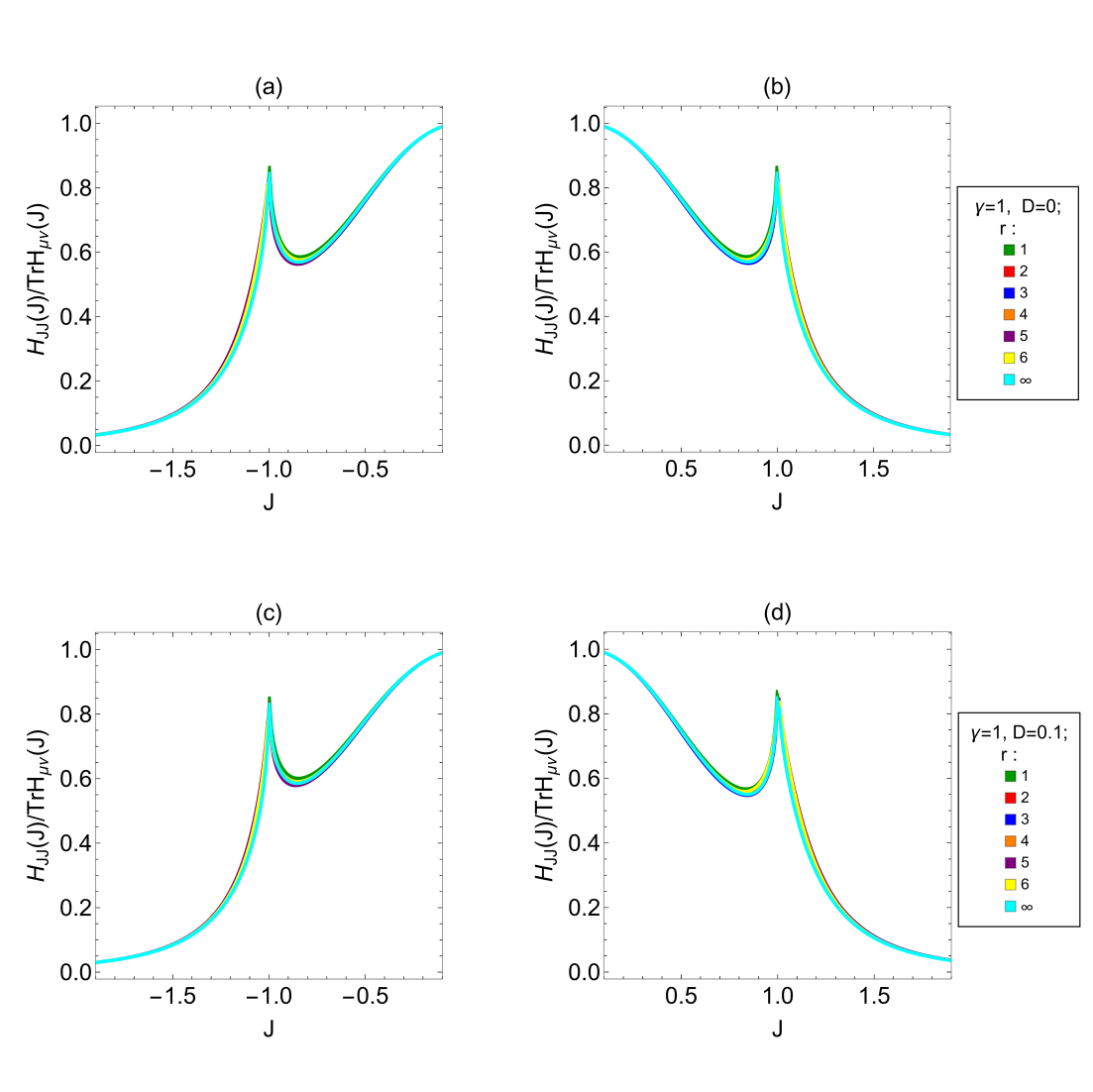}
        \caption{$H_{JJ}/$Tr$[H]$ as $J$ varies.--(a)-(b) Ising model without DM interaction. (c)-(d) Ising model with $D=0.1$.-- Different curves are associated to different distances between the two spins measured $r$. $r=\infty$ correspond to the limit of infinitely distant spins.}
        \label{fig:HJJ_H_Ising}
\end{figure}

Since the sloppiness of the system is large, one may wonder which is the relevant parameter governing the behavior of the system. To this aim it is useful to analyze the ratio between the QFI of, say, the coupling {constant $J$}, and the trace of the entire QFI. In Fig. \ref{fig:HJJ_H_Ising}, we show  the ratio $H_{JJ}/$Tr$[H]$ as a function of $J$ for the different values of $r$, either with or without DM interaction. 

These plots show that $H_{JJ}$ represents the main contribution to the trace in the most part of the $J$ domain. Where it is not, the trace value is very low.  This implies that $J$ carries the most relevant part of the information about the system behavior, and justify our choice to focus our single parameter study on the QFI and the FI associated to $J$. 

The last step of our work is to study the lower bound to precision for the joint estimation of the Hamiltonian parameters. As explained in Section \ref{sec:FI_&_QFI}, this bound is provided by the trace of the inverse of the of Tr$[H^{-1}]$  as a function of $J$ and different $r$, either with and without DM interaction.  

\begin{figure}[!ht]
        \centering
        \includegraphics[width=0.96\columnwidth]{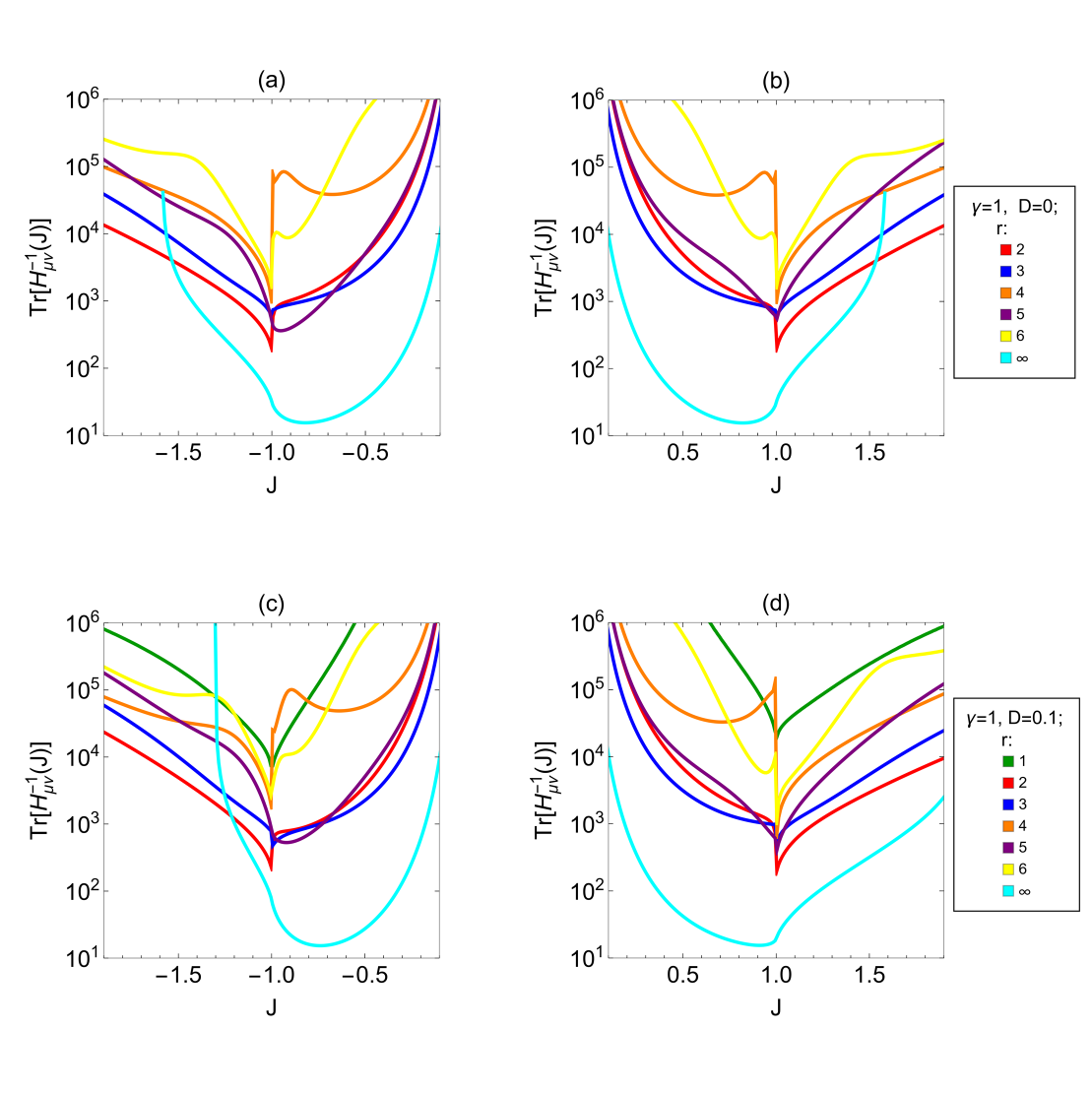}
        \caption{Logplot of Tr$[H^{-1}]$ as a function of $J$.--(a)-(b) Ising model without DM interaction. (c)-(d) Ising model with D=0.1. -- Different curves are associated to different distances between the two spins measured $r$. $r=\infty$ correspond to the limit of infinitely distant spins.}
        \label{fig:Bounds_multiparameter}
\end{figure}

For the Ising model with no DM interaction the bound associated to $r=1$ is not present and the bound associated to $r=\infty$ disappear for $|J| \gtrsim 1.65$. These facts depends on the behavior of the determinant, which is too low in these cases to invert numerically the Fisher matrix. From Fig.\ref{fig:Bounds_multiparameter} we can see how there is no monotonicity respect $r$ either for $D=0$ or $D=0.1$. The most relevant feature in these plots is the order of magnitude for the different neighbors. In the ferromagnetic phase ($|J|<1$) of the Ising model without DM interaction, the difference between the bound for $r=\infty$ and the bounds associated to the other distances oscillates between one or two order of magnitude. In the paramagnetic phase the difference is lower an moving away from the phase transition it decreases up to $|J| \approx 1.6$. After this value, the bound for $r=\infty$ start to cross the the others. For the Ising model with $D=0.1$, the general behavior of the bounds is the same  in ferromagnetic phase and for $J<-1$, but in this region the crossing value is $J \approx -1.2$. For $J>1$ instead, the difference between the bound associated to $r=\infty$ and the bounds associated to the other distances varies from one to two orders of magnitude, as it is in ferromagnetic phase. We also studied the effects of DM interaction up to $D=0.3$ and in this range the bounds for $J>0$ are not significantly affected by $D$. On the other hand for $J<0$, the increase $D$ the makes crossing value to shift to higher values of $J$, whereas the value of the bound for $r=\infty$ at $J=-1$ decreases. 
\section{Conclusions}
\label{sec:Conclusions}
In this paper, we have addressed quantum metrology in critical spin chains with anisotropy and Dzyaloshinskii-Moriya (DM) interactions, and have shown how local and quasi-local measurements may be exploited to characterize global properties 
of the systems. In particular, we have shown that using local measurements collective  phenomena such as different phases of the physical model can be discriminated from the analysis of a single element of the system. This implies that for systems described by the Hamiltonian (\eqref{eq:H_syst}), quantum correlations do not prevent the precise characterization of the system itself by measuring only one of its subparts.

We have also shown that upon measuring just two spins at a given distance, one may exploit correlations to precisely characterize the system, i.e. to estimate the Hamiltonian parameters $J$, $\gamma$ and $D$, and that this gain persists even for infinitely distant spins, where correlations vanish. In particular, we have analytically shown that for two infinitely distant spins, the QFI and the FI for magnetization measurements coincide and are twice the corresponding single spin quantities.

For a general measurement involving two spins, the optimal distance between them depends on all the Hamiltonian parameters $J$, $\gamma$ and $D$ and on the external magnetic field applied to the system. Moreover, this distance is in general different for the QFI and the FI of a magnetization measurements. In other words, the correlations among the spins may have a beneficial or a detrimental role depending on the Hamiltonian parameters and on the distance between the measured spins $r$. 

We have also addressed the joint estimation of all the Hamiltonian parameters, and have shown that it is possible without intrinsic noise, regardless the distance between the measured spins (i.e. the Uhlmann matrix is vanishing $\forall r$). We have then studied the determinant of the QFI matrix in order to quantify the sloppiness of the system, and have shown that it is strongly influenced by the distance between the measured spins. Our results show that for $1 \leq r \leq 6$ the degree of sloppiness of the system is large, except close to the phase transitions of the system. On the other hand, in the limit of infinitely distant spins, there are wide intervals in $J$ for which the sloppiness is low. To conclude the work we have analyzed the lower bounds 
to precision in the multi-parameter case, finding out that the optimal 
distance is $r=\infty$, since the associated bound reaches the lowest values $\forall D$.

\section*{Acknowledgements}
This work was done under the auspices of GNFM-INdAM and have been supported 
by KU through the project C2PS-8474000137. This work has been also partially supported by EU and MIUR through the projects PRIN22-2022T25TR3-RISQUE and 
PRIN22-PNRR-P202222WBL-QWEST.

\appendix

\section{Anisotropic XY model with $\gamma=0.25$}
\label{app:appendix_A}
\begin{figure}[h!]
    \centering
    \includegraphics[width=0.75\columnwidth]{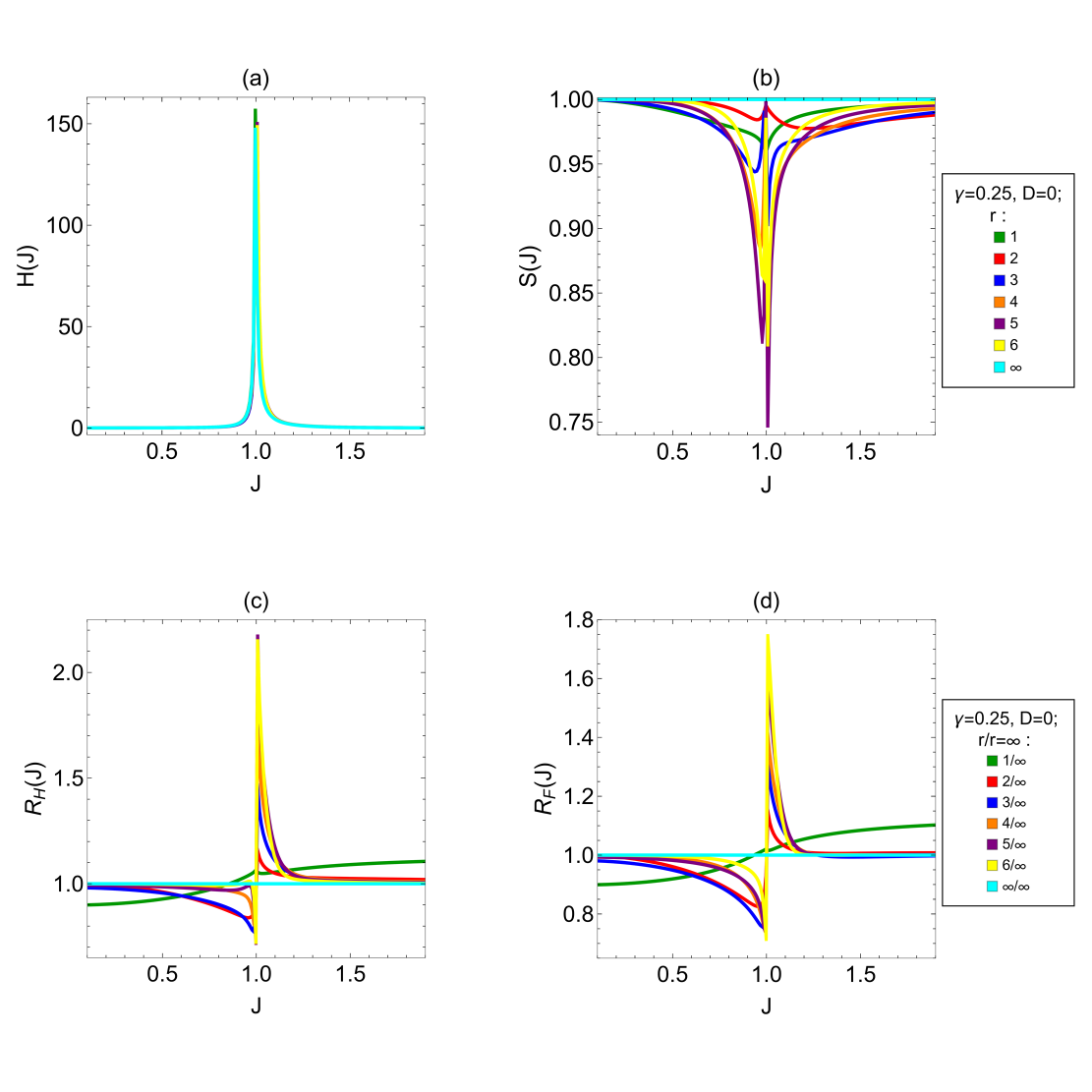}
    \caption{--(a) QFI of coupling constant $J$, $H(J)$. (b) Saturation associated to $J$, $S(J)$. (c)-(d) Ratios between the (Q)FI of the different neighbors and QFI in the limit of infinitely distant neighbors $R_{(H)F}(J)$. -- All the quantities are plotted as $J$ varies, $\gamma=0.25$ and $D=0$. Different curves represent different distances between the two spins measured $r$. $r=\infty$ correspond to the limit of infinitely distant spins.}
    \label{fig:Appendix_D0_g025}
\end{figure}
\begin{figure}[h!]
    \centering
    \includegraphics[width=0.75\columnwidth]{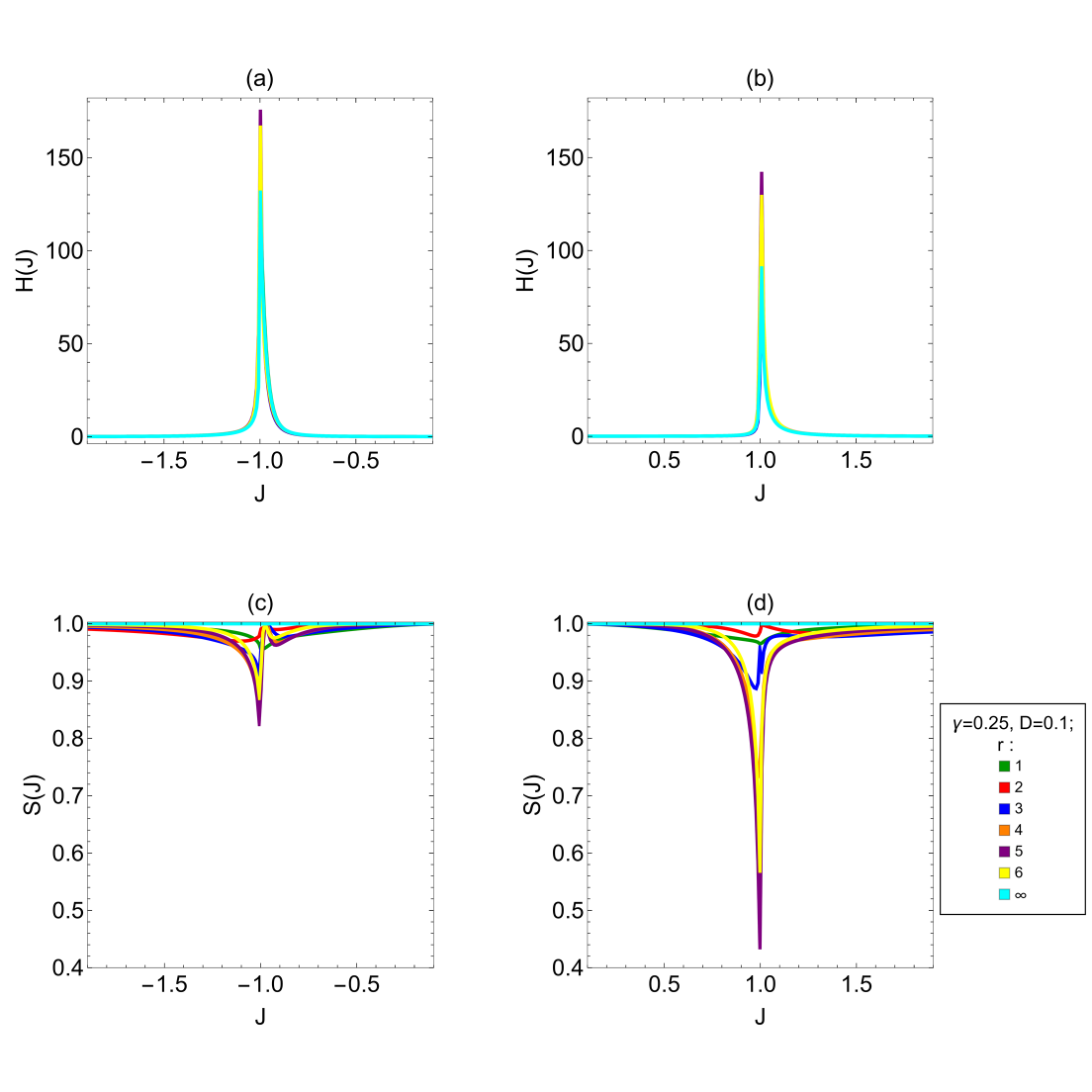}
    \caption{--(a)-(b) QFI of the coupling constant $J$. (c)-(d) Saturation $S(J)$ associated to the coupling constant $J$. In the limit of infinitely distant spins $H(J)=F(J)$. -- All the plots in figure are as $J$ varies, $\gamma=0.25$ and $D=0.1$. The different curves are associated to the different values of the distance between the two spins measured $r$.}
    \label{fig:Appendix_D01_g025_H_S}
\end{figure}
\begin{figure}[h!]
    \centering
    \includegraphics[width=0.75\columnwidth]{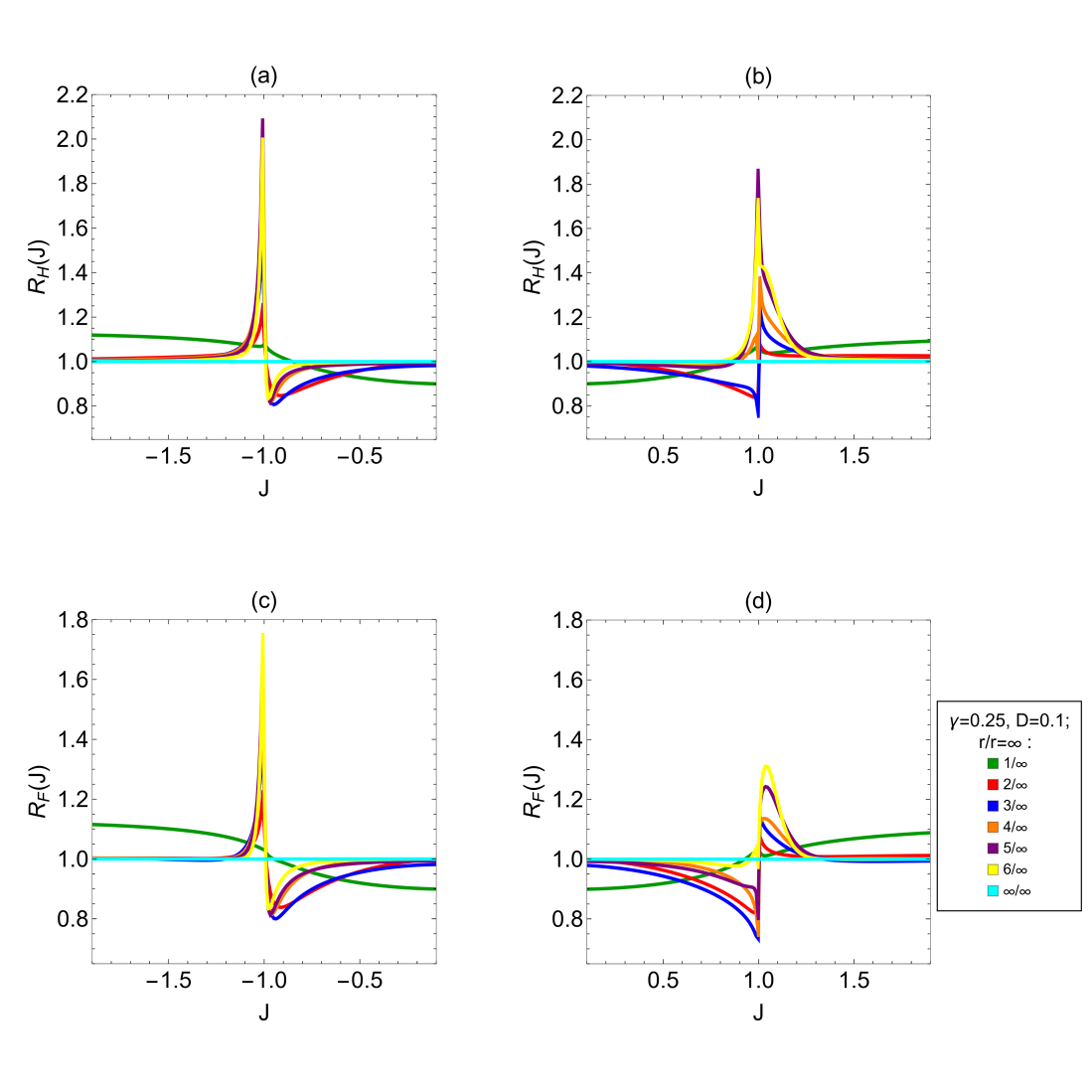}
    \caption{ Ratios between the (Q)FI of different neighbors and (Q)FI in the limit of infinitely distant neighbors, $R_{(H)F}(J)$. -All the quantities are plotted as $J$ varies, $\gamma=0.25$ and $D=0.1$. Different curves represent different distances between the two spins measured $r$. $r=\infty$ correspond to the limit of infinitely distant spins.}
    \label{fig:Appendix_D01_g025_RH_RF}
\end{figure}

To support what we state in Sec. \ref{sec:Quasi_local_measurements}, we provide further examples of the behavior of the (Q)FI associated to different values of $\gamma$ and $D$. We report the results for the quasi-local estimation $J$, for $\gamma=0.25$ without DM interaction, $D=0$, and with DM factor $D=0.1$. The main characteristics of the (Q)FI remain the same independently on the values of $\gamma$ and $D$, promoting what we observed in Sec.\ref{sec:Quasi_local_measurements} to more general considerations about our anisotropic model. A magnetization measurement is then capable to extract a consistent part of the information present in the system, for all the neighbors studied. Nevertheless, there are still differences with respect to the Ising model, since the (Q)FI assumes higher values with $\gamma=0.25$, with respect to $\gamma=1$ (see Fig.\ref{fig:Appendix_D0_g025}, \ref{fig:Appendix_D01_g025_H_S}, and \ref{fig:Appendix_D01_g025_RH_RF}).

\clearpage

\section{Uhlmann matrix for real X-states}
\label{app:appendix_B}

 {In this appendix,  we prove that the Uhlmann matrix associated with a real X-state is always a null matrix. To do so, we start from the notions about the X-states contained in \cite{maroufi2021analytical}}. A general X-state $\rho(\lambda)$ {has a matrix representation in the form}

    \begin{equation}
        \rho(\lambda) =
        \begin{pmatrix}
				\rho_{11} & 0 & 0 & \rho_{14}  \\
				0 & \rho_{22} & \rho_{23} & 0  \\
				0 & \rho_{32} & \rho_{33} & 0  \\
				\rho_{41} & 0 & 0 & \rho_{44}
		\end{pmatrix} \ .
	\end{equation}
It can be decomposed as $\rho(\lambda) = \rho_1(\lambda) + \rho_2(\lambda)$. Where 
    \begin{equation}
		\rho_1(\lambda) = \frac{1}{2} \sum_{ \alpha = 0}^3 \ \omega_\alpha \eta_\alpha \ , \ \ \ \ \ \ \ \rho_2(\lambda) = \frac{1}{2} \sum_{ \alpha = 0}^3 \ \tilde{\omega}_\alpha \tilde{\eta}_\alpha \ .
	\end{equation}
In these expressions $\eta_\alpha$ and $\tilde{\eta}_\alpha$ are defined by
	\begin{equation}
		\begin{split}
			&\eta_0 = |00 \rangle \langle 00| + |11 \rangle \langle 11|,  \\
			&\eta_1 = |00 \rangle \langle 11| + |11 \rangle \langle 00|,  \\
			&\eta_2 = i|11 \rangle \langle 00| - i|00 \rangle \langle 11|,  \\
			&\eta_3 = |00 \rangle \langle 00| - |11 \rangle \langle 11|, \\
            &\tilde{\eta}_0 = |01 \rangle \langle 01| + |10 \rangle \langle 10|, \\
            &\tilde{\eta}_1 = |01 \rangle \langle 10| + |10 \rangle \langle 01|, \\
            &\tilde{\eta}_2 = i|10 \rangle \langle 01| - i|01 \rangle \langle 10|, \\
            &\tilde{\eta}_3 = |01 \rangle \langle 01| - |10 \rangle \langle 10| 
		\end{split}
	\end{equation}
And the $\omega_\alpha$ and $\tilde{\omega}_\alpha$ by
\begin{equation}
    \begin{split}
        &\omega_0 = ( \rho_{11} + \rho_{44} ), \ \ \ \ \omega_1 = ( \rho_{14} + \rho_{41}), \\
        &\omega_2 = i( \rho_{14} - \rho_{41} ), \ \ \ \ \omega_3 = ( \rho_{11} - \rho_{44} ), \\
        &\tilde{\omega}_0 = ( \rho_{22} + \rho_{33} ), \ \ \ \ \tilde{\omega}_1 = ( \rho_{23} + \rho_{32}), \\
        &\tilde{\omega}_2 = i( \rho_{23} - \rho_{32} ), \ \ \ \ \tilde{\omega}_3 = ( \rho_{22} - \rho_{33} ).
    \end{split}
\end{equation}
If $\rho(\lambda)$ is real (as in our case), we have $\rho_{ij}=\rho_{ji}^{*}=\rho_{ji}$, so
\begin{equation}
    \begin{split}
        &\omega_1 = 2 \rho_{14}=2b_{-}, \ \ \ \
        \omega_2 = 0, \\
        &\tilde{\omega}_1 = 2 \rho_{23}=2b_{+}, \ \ \ \
        \tilde{\omega}_2 = 0.
    \end{split}
\end{equation}
Also the Symmetric Logarithmic Derivative associated to $\rho(\lambda)$ can be decomposed as $\mathcal{L}_{\lambda} = \mathcal{L}_{\lambda}^{(1)}+ \mathcal{L}_{\lambda}^{(2)} $, where $\mathcal{L}_{\lambda}^{(1)}$ is the SLD associated to $\rho_1(\lambda)$ and $\mathcal{L}_{\lambda}^{(2)}$ the SLD associated to $\rho_2(\lambda)$. $\mathcal{L}_{\lambda}^{(1)}$ and $\mathcal{L}_{\lambda}^{(2)}$ can be written as
\begin{equation}
\begin{split}
     &\mathcal{L}_{\lambda}^{(1)} =\sum_{ \alpha = 0}^3 \ f_\alpha \eta_\alpha \ ,   \\ 
     &\mathcal{L}_{\lambda}^{(2)} =\sum_{ \alpha = 0}^3 \ \tilde{f}_\alpha \tilde{\eta}_\alpha \ .
\end{split}
\end{equation}
Where
\begin{equation}
    	\begin{split}
    		& f_0 = \frac{\omega_0(\partial_\lambda \omega_0)- \sum_i \omega_i (\partial_\lambda \omega_i)}{\omega_0^2 - \sum_i \omega_i^2}, \\
            & f_i = \frac{\partial_\lambda \omega_i - f_0 \omega_i}{\omega_0}, \\
            & \tilde{f}_0 = \frac{\tilde{\omega}_0(\partial_\lambda \tilde{\omega}_0)- \sum_i \tilde{\omega}_i (\partial_\lambda \tilde{\omega}_i)}{\tilde{\omega}_0^2 - \sum_i \tilde{\omega}_i^2}, \\
            & \tilde{f}_i = \frac{\partial_\lambda \tilde{\omega}_i - \tilde{b}_0 \tilde{\omega}_i}{\tilde{\omega}_0}, \ \ i=1,2,3.
    	\end{split}
    \end{equation}
When $\rho(\lambda)$ is real we have $f_2=\tilde{f}_2=0$, since $\omega_2=\tilde{\omega}_2=0$, and therefore the SLD $\mathcal{L}(\lambda) = \mathcal{L}_{\lambda}^{(1)} + \mathcal{L}_{\lambda}^{(2)}$ is symmetric and has a X structure. This is the main point for the following. Up to now we used the formalism of the single parameter estimation for simplicity ($\lambda$ is scalar). From now on we use the formalism of the multi-parameter estimation, so now $\boldsymbol{\lambda}$ is the vector of the parameters to be estimated. Now let's look for the Uhlmann matrix (Eq.\eqref{eq:general_Uhlmann_matrix}). When $\rho(\lambda)$ is real and has an X structure, we can write the SLD associate to a generic parameter $\mu\in\boldsymbol{\lambda}$ and the SLD associate to a generic parameter $\nu\in\boldsymbol{\lambda}$ as
\begin{equation}
    \begin{split}
        &\mathcal{L}_{\mu} =\begin{pmatrix}
            \mu_{11} & 0 & 0 & \mu_{14}  \\
				0 & \mu_{22} & \mu_{23} & 0  \\
				0 & \mu_{23} & \mu_{33} & 0  \\
				\mu_{14} & 0 & 0 & \mu_{44}
        \end{pmatrix}, \\ 
        &\mathcal{L}_{\nu} = \begin{pmatrix}
            \nu_{11} & 0 & 0 & \nu_{14}  \\
				0 & \nu_{22} & \nu_{23} & 0  \\
				0 & \nu_{23} & \nu_{33} & 0  \\
				\nu_{14} & 0 & 0 & \nu_{44}
        \end{pmatrix}.
    \end{split}
\end{equation}
Now to find the elements of the Uhlmann matrix we can just apply 
\begin{equation}
    \label{eq:Uhlmann_matrix_AppB}
        {\boldsymbol{U}}_{\mu\nu}\boldsymbol{(\lambda)} = Tr\left[\rho_{\boldsymbol{\lambda}}\frac{\mathcal{L}_{\mu}\mathcal{L}_{\nu}-\mathcal{L}_{\nu}\mathcal{L}_{\mu}}{2}\right] = \hbox{Tr}\left[M\right] \ .
\end{equation}
Performing the algebra we find the diagonal elements $M_{ii}$ of the matrix inside the trace in the equation above (Eq.\eqref{eq:Uhlmann_matrix_AppB}). These are
\begin{equation}
    \begin{split}
        &M_{11}=\frac{1}{2}((-\mu_{11}+\mu{44})\nu_{14}+\mu_{14}(\nu_{11}-\nu_{44}))b_{-}, \\
        &M_{22}=\frac{1}{2}((-\mu_{22}+\mu{33})\nu_{23}+\mu_{23}(\nu_{22}-\nu_{33}))b_{+}, \\
        &M_{33}=\frac{1}{2}((\mu_{22}-\mu{33})\nu_{23}+\mu_{23}(-\nu_{22}+\nu_{33}))b_{+}, \\
        &M_{44}=\frac{1}{2}((\mu_{11}-\mu{44})\nu_{14}+\mu_{14}(-\nu_{11}+\nu_{44}))b_{-},
    \end{split}
\end{equation}
so $M_{11}=-M_{44}$ and $M_{22}=-M_{33}$. For this reason taking trace of $M$ we obtain
\begin{equation}
    {\boldsymbol{U}}_{\mu\nu}\boldsymbol{(\lambda)} = 0, \ \ \ \ \forall \ \mu,\nu \in \boldsymbol{\lambda} .
\end{equation}
This means that for the systems described by a real X-state the Uhlmann matrix is always null and the Symmetric Logarithmic Derivatives commutes weakly among each other. For this reason all the systems of this kind are classified as \textit{asymptotically classical systems} \cite{ALBARELLI2020126311}.

\bibliography{masp.bib}

\begin{thebibliography}{59}%
\makeatletter
\providecommand \@ifxundefined [1]{%
 \@ifx{#1\undefined}
}%
\providecommand \@ifnum [1]{%
 \ifnum #1\expandafter \@firstoftwo
 \else \expandafter \@secondoftwo
 \fi
}%
\providecommand \@ifx [1]{%
 \ifx #1\expandafter \@firstoftwo
 \else \expandafter \@secondoftwo
 \fi
}%
\providecommand \natexlab [1]{#1}%
\providecommand \enquote  [1]{``#1''}%
\providecommand \bibnamefont  [1]{#1}%
\providecommand \bibfnamefont [1]{#1}%
\providecommand \citenamefont [1]{#1}%
\providecommand \href@noop [0]{\@secondoftwo}%
\providecommand \href [0]{\begingroup \@sanitize@url \@href}%
\providecommand \@href[1]{\@@startlink{#1}\@@href}%
\providecommand \@@href[1]{\endgroup#1\@@endlink}%
\providecommand \@sanitize@url [0]{\catcode `\\12\catcode `\$12\catcode
  `\&12\catcode `\#12\catcode `\^12\catcode `\_12\catcode `\%12\relax}%
\providecommand \@@startlink[1]{}%
\providecommand \@@endlink[0]{}%
\providecommand \url  [0]{\begingroup\@sanitize@url \@url }%
\providecommand \@url [1]{\endgroup\@href {#1}{\urlprefix }}%
\providecommand \urlprefix  [0]{URL }%
\providecommand \Eprint [0]{\href }%
\providecommand \doibase [0]{https://doi.org/}%
\providecommand \selectlanguage [0]{\@gobble}%
\providecommand \bibinfo  [0]{\@secondoftwo}%
\providecommand \bibfield  [0]{\@secondoftwo}%
\providecommand \translation [1]{[#1]}%
\providecommand \BibitemOpen [0]{}%
\providecommand \bibitemStop [0]{}%
\providecommand \bibitemNoStop [0]{.\EOS\space}%
\providecommand \EOS [0]{\spacefactor3000\relax}%
\providecommand \BibitemShut  [1]{\csname bibitem#1\endcsname}%
\let\auto@bib@innerbib\@empty
\bibitem [{\citenamefont {Sachdev}(2000)}]{Sachdev_2000}%
  \BibitemOpen
  \bibfield  {author} {\bibinfo {author} {\bibfnamefont {S.}~\bibnamefont
  {Sachdev}},\ }\href@noop {} {\emph {\bibinfo {title} {Quantum Phase
  Transitions}}}\ (\bibinfo  {publisher} {Cambridge University Press},\
  \bibinfo {year} {2000})\BibitemShut {NoStop}%
\bibitem [{\citenamefont {Zanardi}\ \emph {et~al.}(2008)\citenamefont
  {Zanardi}, \citenamefont {Paris},\ and\ \citenamefont
  {Venuti}}]{zanardi2008quantum}%
  \BibitemOpen
  \bibfield  {author} {\bibinfo {author} {\bibfnamefont {P.}~\bibnamefont
  {Zanardi}}, \bibinfo {author} {\bibfnamefont {M.~G.}\ \bibnamefont {Paris}},\
  and\ \bibinfo {author} {\bibfnamefont {L.~C.}\ \bibnamefont {Venuti}},\
  }\bibfield  {title} {\bibinfo {title} {Quantum criticality as a resource for
  quantum estimation},\ }\href@noop {} {\bibfield  {journal} {\bibinfo
  {journal} {Physical Review A}\ }\textbf {\bibinfo {volume} {78}},\ \bibinfo
  {pages} {042105} (\bibinfo {year} {2008})}\BibitemShut {NoStop}%
\bibitem [{\citenamefont {Macieszczak}\ \emph {et~al.}(2016)\citenamefont
  {Macieszczak}, \citenamefont {Gu{\c{t}}{\u{a}}}, \citenamefont {Lesanovsky},\
  and\ \citenamefont {Garrahan}}]{macieszczak2016dynamical}%
  \BibitemOpen
  \bibfield  {author} {\bibinfo {author} {\bibfnamefont {K.}~\bibnamefont
  {Macieszczak}}, \bibinfo {author} {\bibfnamefont {M.}~\bibnamefont
  {Gu{\c{t}}{\u{a}}}}, \bibinfo {author} {\bibfnamefont {I.}~\bibnamefont
  {Lesanovsky}},\ and\ \bibinfo {author} {\bibfnamefont {J.~P.}\ \bibnamefont
  {Garrahan}},\ }\bibfield  {title} {\bibinfo {title} {Dynamical phase
  transitions as a resource for quantum enhanced metrology},\ }\href@noop {}
  {\bibfield  {journal} {\bibinfo  {journal} {Physical Review A}\ }\textbf
  {\bibinfo {volume} {93}},\ \bibinfo {pages} {022103} (\bibinfo {year}
  {2016})}\BibitemShut {NoStop}%
\bibitem [{\citenamefont {Song}\ \emph {et~al.}(2017)\citenamefont {Song},
  \citenamefont {Luo},\ and\ \citenamefont {Fu}}]{song2017quantum}%
  \BibitemOpen
  \bibfield  {author} {\bibinfo {author} {\bibfnamefont {H.}~\bibnamefont
  {Song}}, \bibinfo {author} {\bibfnamefont {S.}~\bibnamefont {Luo}},\ and\
  \bibinfo {author} {\bibfnamefont {S.}~\bibnamefont {Fu}},\ }\bibfield
  {title} {\bibinfo {title} {Quantum criticality from \uppercase {F}isher
  information},\ }\href@noop {} {\bibfield  {journal} {\bibinfo  {journal}
  {Quantum Information Processing}\ }\textbf {\bibinfo {volume} {16}},\
  \bibinfo {pages} {1} (\bibinfo {year} {2017})}\BibitemShut {NoStop}%
\bibitem [{\citenamefont {Garbe}\ \emph {et~al.}(2020)\citenamefont {Garbe},
  \citenamefont {Bina}, \citenamefont {Keller}, \citenamefont {Paris},\ and\
  \citenamefont {Felicetti}}]{garbe2020critical}%
  \BibitemOpen
  \bibfield  {author} {\bibinfo {author} {\bibfnamefont {L.}~\bibnamefont
  {Garbe}}, \bibinfo {author} {\bibfnamefont {M.}~\bibnamefont {Bina}},
  \bibinfo {author} {\bibfnamefont {A.}~\bibnamefont {Keller}}, \bibinfo
  {author} {\bibfnamefont {M.~G.}\ \bibnamefont {Paris}},\ and\ \bibinfo
  {author} {\bibfnamefont {S.}~\bibnamefont {Felicetti}},\ }\bibfield  {title}
  {\bibinfo {title} {Critical quantum metrology with a finite-component quantum
  phase transition},\ }\href@noop {} {\bibfield  {journal} {\bibinfo  {journal}
  {Physical review letters}\ }\textbf {\bibinfo {volume} {124}},\ \bibinfo
  {pages} {120504} (\bibinfo {year} {2020})}\BibitemShut {NoStop}%
\bibitem [{\citenamefont {Jafari}\ \emph {et~al.}(2008)\citenamefont {Jafari},
  \citenamefont {Kargarian}, \citenamefont {Langari},\ and\ \citenamefont
  {Siahatgar}}]{jafari2008phase}%
  \BibitemOpen
  \bibfield  {author} {\bibinfo {author} {\bibfnamefont {R.}~\bibnamefont
  {Jafari}}, \bibinfo {author} {\bibfnamefont {M.}~\bibnamefont {Kargarian}},
  \bibinfo {author} {\bibfnamefont {A.}~\bibnamefont {Langari}},\ and\ \bibinfo
  {author} {\bibfnamefont {M.}~\bibnamefont {Siahatgar}},\ }\bibfield  {title}
  {\bibinfo {title} {Phase diagram and entanglement of the ising model with
  dzyaloshinskii-moriya interaction},\ }\href@noop {} {\bibfield  {journal}
  {\bibinfo  {journal} {Physical Review B}\ }\textbf {\bibinfo {volume} {78}},\
  \bibinfo {pages} {214414} (\bibinfo {year} {2008})}\BibitemShut {NoStop}%
\bibitem [{\citenamefont {Messio}\ \emph {et~al.}(2010)\citenamefont {Messio},
  \citenamefont {Cepas},\ and\ \citenamefont
  {Lhuillier}}]{messio2010schwinger}%
  \BibitemOpen
  \bibfield  {author} {\bibinfo {author} {\bibfnamefont {L.}~\bibnamefont
  {Messio}}, \bibinfo {author} {\bibfnamefont {O.}~\bibnamefont {Cepas}},\ and\
  \bibinfo {author} {\bibfnamefont {C.}~\bibnamefont {Lhuillier}},\ }\bibfield
  {title} {\bibinfo {title} {Schwinger-boson approach to the kagome
  antiferromagnet with dzyaloshinskii-moriya interactions: Phase diagram and
  dynamical structure factors},\ }\href@noop {} {\bibfield  {journal} {\bibinfo
   {journal} {Physical Review B}\ }\textbf {\bibinfo {volume} {81}},\ \bibinfo
  {pages} {064428} (\bibinfo {year} {2010})}\BibitemShut {NoStop}%
\bibitem [{\citenamefont {Garate}\ and\ \citenamefont
  {Affleck}(2010)}]{garate2010interplay}%
  \BibitemOpen
  \bibfield  {author} {\bibinfo {author} {\bibfnamefont {I.}~\bibnamefont
  {Garate}}\ and\ \bibinfo {author} {\bibfnamefont {I.}~\bibnamefont
  {Affleck}},\ }\bibfield  {title} {\bibinfo {title} {Interplay between
  symmetric exchange anisotropy, uniform dzyaloshinskii-moriya interaction, and
  magnetic fields in the phase diagram of quantum magnets and
  superconductors},\ }\href@noop {} {\bibfield  {journal} {\bibinfo  {journal}
  {Physical Review B}\ }\textbf {\bibinfo {volume} {81}},\ \bibinfo {pages}
  {144419} (\bibinfo {year} {2010})}\BibitemShut {NoStop}%
\bibitem [{\citenamefont {Parente}\ \emph {et~al.}(2015)\citenamefont
  {Parente}, \citenamefont {Pacobahyba}, \citenamefont {Ara{\'u}jo},
  \citenamefont {Neto},\ and\ \citenamefont {de~Sousa}}]{parente2015anomaly}%
  \BibitemOpen
  \bibfield  {author} {\bibinfo {author} {\bibfnamefont {W.~E.}\ \bibnamefont
  {Parente}}, \bibinfo {author} {\bibfnamefont {J.}~\bibnamefont {Pacobahyba}},
  \bibinfo {author} {\bibfnamefont {I.~G.}\ \bibnamefont {Ara{\'u}jo}},
  \bibinfo {author} {\bibfnamefont {M.~A.}\ \bibnamefont {Neto}},\ and\
  \bibinfo {author} {\bibfnamefont {J.~R.}\ \bibnamefont {de~Sousa}},\
  }\bibfield  {title} {\bibinfo {title} {Anomaly in the phase diagram of the
  spin quantum 1/2 anisotropic heisenberg antiferromagnet model with
  dzyaloshinskii--moriya interaction: A low temperature analysis},\ }\href@noop
  {} {\bibfield  {journal} {\bibinfo  {journal} {Physica E: Low-dimensional
  Systems and Nanostructures}\ }\textbf {\bibinfo {volume} {74}},\ \bibinfo
  {pages} {287} (\bibinfo {year} {2015})}\BibitemShut {NoStop}%
\bibitem [{\citenamefont {Marzolino}\ and\ \citenamefont
  {Prosen}(2017)}]{marzolino2017fisher}%
  \BibitemOpen
  \bibfield  {author} {\bibinfo {author} {\bibfnamefont {U.}~\bibnamefont
  {Marzolino}}\ and\ \bibinfo {author} {\bibfnamefont {T.}~\bibnamefont
  {Prosen}},\ }\bibfield  {title} {\bibinfo {title} {Fisher information
  approach to nonequilibrium phase transitions in a quantum xxz spin chain with
  boundary noise},\ }\href@noop {} {\bibfield  {journal} {\bibinfo  {journal}
  {Physical Review B}\ }\textbf {\bibinfo {volume} {96}},\ \bibinfo {pages}
  {104402} (\bibinfo {year} {2017})}\BibitemShut {NoStop}%
\bibitem [{\citenamefont {Jin}\ and\ \citenamefont
  {Starykh}(2017)}]{jin2017phase}%
  \BibitemOpen
  \bibfield  {author} {\bibinfo {author} {\bibfnamefont {W.}~\bibnamefont
  {Jin}}\ and\ \bibinfo {author} {\bibfnamefont {O.~A.}\ \bibnamefont
  {Starykh}},\ }\bibfield  {title} {\bibinfo {title} {Phase diagram of weakly
  coupled heisenberg spin chains subject to a uniform dzyaloshinskii-moriya
  interaction},\ }\href@noop {} {\bibfield  {journal} {\bibinfo  {journal}
  {Physical Review B}\ }\textbf {\bibinfo {volume} {95}},\ \bibinfo {pages}
  {214404} (\bibinfo {year} {2017})}\BibitemShut {NoStop}%
\bibitem [{\citenamefont {Yi}\ \emph {et~al.}(2019)\citenamefont {Yi},
  \citenamefont {You}, \citenamefont {Wu},\ and\ \citenamefont
  {Ole{\'s}}}]{yi2019criticality}%
  \BibitemOpen
  \bibfield  {author} {\bibinfo {author} {\bibfnamefont {T.-C.}\ \bibnamefont
  {Yi}}, \bibinfo {author} {\bibfnamefont {W.-L.}\ \bibnamefont {You}},
  \bibinfo {author} {\bibfnamefont {N.}~\bibnamefont {Wu}},\ and\ \bibinfo
  {author} {\bibfnamefont {A.~M.}\ \bibnamefont {Ole{\'s}}},\ }\bibfield
  {title} {\bibinfo {title} {Criticality and factorization in the heisenberg
  chain with dzyaloshinskii-moriya interaction},\ }\href@noop {} {\bibfield
  {journal} {\bibinfo  {journal} {Physical Review B}\ }\textbf {\bibinfo
  {volume} {100}},\ \bibinfo {pages} {024423} (\bibinfo {year}
  {2019})}\BibitemShut {NoStop}%
\bibitem [{\citenamefont {Thakur}\ and\ \citenamefont
  {Durganandini}(2020)}]{thakur2020factorization}%
  \BibitemOpen
  \bibfield  {author} {\bibinfo {author} {\bibfnamefont {P.}~\bibnamefont
  {Thakur}}\ and\ \bibinfo {author} {\bibfnamefont {P.}~\bibnamefont
  {Durganandini}},\ }\bibfield  {title} {\bibinfo {title} {Factorization,
  coherence, and asymmetry in the heisenberg spin-1 2 xxz chain with
  dzyaloshinskii-moriya interaction in transverse magnetic field},\ }\href@noop
  {} {\bibfield  {journal} {\bibinfo  {journal} {Physical Review B}\ }\textbf
  {\bibinfo {volume} {102}},\ \bibinfo {pages} {064409} (\bibinfo {year}
  {2020})}\BibitemShut {NoStop}%
\bibitem [{\citenamefont {Japaridze}\ \emph {et~al.}(2021)\citenamefont
  {Japaridze}, \citenamefont {Cheraghi},\ and\ \citenamefont
  {Mahdavifar}}]{japaridze2021magnetic}%
  \BibitemOpen
  \bibfield  {author} {\bibinfo {author} {\bibfnamefont {G.}~\bibnamefont
  {Japaridze}}, \bibinfo {author} {\bibfnamefont {H.}~\bibnamefont
  {Cheraghi}},\ and\ \bibinfo {author} {\bibfnamefont {S.}~\bibnamefont
  {Mahdavifar}},\ }\bibfield  {title} {\bibinfo {title} {Magnetic phase diagram
  of a spin-1/2 x x z chain with modulated dzyaloshinskii-moriya interaction},\
  }\href@noop {} {\bibfield  {journal} {\bibinfo  {journal} {Physical Review
  E}\ }\textbf {\bibinfo {volume} {104}},\ \bibinfo {pages} {014134} (\bibinfo
  {year} {2021})}\BibitemShut {NoStop}%
\bibitem [{\citenamefont {Fumani}\ \emph {et~al.}(2021)\citenamefont {Fumani},
  \citenamefont {Beradze}, \citenamefont {Nemati}, \citenamefont {Mahdavifar},\
  and\ \citenamefont {Japaridze}}]{fumani2021quantum}%
  \BibitemOpen
  \bibfield  {author} {\bibinfo {author} {\bibfnamefont {F.~K.}\ \bibnamefont
  {Fumani}}, \bibinfo {author} {\bibfnamefont {B.}~\bibnamefont {Beradze}},
  \bibinfo {author} {\bibfnamefont {S.}~\bibnamefont {Nemati}}, \bibinfo
  {author} {\bibfnamefont {S.}~\bibnamefont {Mahdavifar}},\ and\ \bibinfo
  {author} {\bibfnamefont {G.}~\bibnamefont {Japaridze}},\ }\bibfield  {title}
  {\bibinfo {title} {Quantum correlations in the spin-1/2 heisenberg xxz chain
  with modulated dzyaloshinskii-moriya interaction},\ }\href@noop {} {\bibfield
   {journal} {\bibinfo  {journal} {Journal of Magnetism and Magnetic
  Materials}\ }\textbf {\bibinfo {volume} {518}},\ \bibinfo {pages} {167411}
  (\bibinfo {year} {2021})}\BibitemShut {NoStop}%
\bibitem [{\citenamefont {Cavazzoni}\ \emph {et~al.}(2024)\citenamefont
  {Cavazzoni}, \citenamefont {Adani}, \citenamefont {Bordone},\ and\
  \citenamefont {Paris}}]{cavazzoni24}%
  \BibitemOpen
  \bibfield  {author} {\bibinfo {author} {\bibfnamefont {S.}~\bibnamefont
  {Cavazzoni}}, \bibinfo {author} {\bibfnamefont {M.}~\bibnamefont {Adani}},
  \bibinfo {author} {\bibfnamefont {P.}~\bibnamefont {Bordone}},\ and\ \bibinfo
  {author} {\bibfnamefont {M.~G.~A.}\ \bibnamefont {Paris}},\ }\bibfield
  {title} {\bibinfo {title} {Characterization of partially accessible
  anisotropic spin chains in the presence of anti-symmetric exchange},\ }\href
  {https://doi.org/10.1088/1367-2630/ad48ae} {\bibfield  {journal} {\bibinfo
  {journal} {New Journal of Physics}\ }\textbf {\bibinfo {volume} {26}},\
  \bibinfo {pages} {053024} (\bibinfo {year} {2024})}\BibitemShut {NoStop}%
\bibitem [{\citenamefont {Mishra}\ and\ \citenamefont {Bayat}(2021)}]{pa1}%
  \BibitemOpen
  \bibfield  {author} {\bibinfo {author} {\bibfnamefont {U.}~\bibnamefont
  {Mishra}}\ and\ \bibinfo {author} {\bibfnamefont {A.}~\bibnamefont {Bayat}},\
  }\bibfield  {title} {\bibinfo {title} {Driving enhanced quantum sensing in
  partially accessible many-body systems},\ }\href
  {https://doi.org/10.1103/PhysRevLett.127.080504} {\bibfield  {journal}
  {\bibinfo  {journal} {Phys. Rev. Lett.}\ }\textbf {\bibinfo {volume} {127}},\
  \bibinfo {pages} {080504} (\bibinfo {year} {2021})}\BibitemShut {NoStop}%
\bibitem [{\citenamefont {Montenegro}\ \emph {et~al.}(2022)\citenamefont
  {Montenegro}, \citenamefont {Genoni}, \citenamefont {Bayat},\ and\
  \citenamefont {Paris}}]{pa2}%
  \BibitemOpen
  \bibfield  {author} {\bibinfo {author} {\bibfnamefont {V.}~\bibnamefont
  {Montenegro}}, \bibinfo {author} {\bibfnamefont {M.~G.}\ \bibnamefont
  {Genoni}}, \bibinfo {author} {\bibfnamefont {A.}~\bibnamefont {Bayat}},\ and\
  \bibinfo {author} {\bibfnamefont {M.~G.~A.}\ \bibnamefont {Paris}},\
  }\bibfield  {title} {\bibinfo {title} {Probing of nonlinear hybrid
  optomechanical systems via partial accessibility},\ }\href
  {https://doi.org/10.1103/PhysRevResearch.4.033036} {\bibfield  {journal}
  {\bibinfo  {journal} {Phys. Rev. Res.}\ }\textbf {\bibinfo {volume} {4}},\
  \bibinfo {pages} {033036} (\bibinfo {year} {2022})}\BibitemShut {NoStop}%
\bibitem [{\citenamefont {Mishra}\ and\ \citenamefont {Bayat}(2022)}]{pa3}%
  \BibitemOpen
  \bibfield  {author} {\bibinfo {author} {\bibfnamefont {U.}~\bibnamefont
  {Mishra}}\ and\ \bibinfo {author} {\bibfnamefont {A.}~\bibnamefont {Bayat}},\
  }\bibfield  {title} {\bibinfo {title} {Integrable quantum many-body sensors
  for ac field sensing},\ }\href {https://doi.org/10.1038/s41598-022-17381-y}
  {\bibfield  {journal} {\bibinfo  {journal} {Scientific Reports}\ }\textbf
  {\bibinfo {volume} {12}},\ \bibinfo {pages} {14760} (\bibinfo {year}
  {2022})}\BibitemShut {NoStop}%
\bibitem [{\citenamefont {Koretsune}\ \emph {et~al.}(2018)\citenamefont
  {Koretsune}, \citenamefont {Kikuchi},\ and\ \citenamefont
  {Arita}}]{koretsune2018first}%
  \BibitemOpen
  \bibfield  {author} {\bibinfo {author} {\bibfnamefont {T.}~\bibnamefont
  {Koretsune}}, \bibinfo {author} {\bibfnamefont {T.}~\bibnamefont {Kikuchi}},\
  and\ \bibinfo {author} {\bibfnamefont {R.}~\bibnamefont {Arita}},\ }\bibfield
   {title} {\bibinfo {title} {First-principles evaluation of the
  dzyaloshinskii--moriya interaction},\ }\href@noop {} {\bibfield  {journal}
  {\bibinfo  {journal} {journal of the physical society of japan}\ }\textbf
  {\bibinfo {volume} {87}},\ \bibinfo {pages} {041011} (\bibinfo {year}
  {2018})}\BibitemShut {NoStop}%
\bibitem [{\citenamefont {Cardias}\ \emph {et~al.}(2020)\citenamefont
  {Cardias}, \citenamefont {Szilva}, \citenamefont {Bezerra-Neto},
  \citenamefont {Ribeiro}, \citenamefont {Bergman}, \citenamefont {Kvashnin},
  \citenamefont {Fransson}, \citenamefont {Klautau}, \citenamefont {Eriksson},\
  and\ \citenamefont {Nordstr{\"o}m}}]{cardias2020first}%
  \BibitemOpen
  \bibfield  {author} {\bibinfo {author} {\bibfnamefont {R.}~\bibnamefont
  {Cardias}}, \bibinfo {author} {\bibfnamefont {A.}~\bibnamefont {Szilva}},
  \bibinfo {author} {\bibfnamefont {M.}~\bibnamefont {Bezerra-Neto}}, \bibinfo
  {author} {\bibfnamefont {M.}~\bibnamefont {Ribeiro}}, \bibinfo {author}
  {\bibfnamefont {A.}~\bibnamefont {Bergman}}, \bibinfo {author} {\bibfnamefont
  {Y.~O.}\ \bibnamefont {Kvashnin}}, \bibinfo {author} {\bibfnamefont
  {J.}~\bibnamefont {Fransson}}, \bibinfo {author} {\bibfnamefont
  {A.}~\bibnamefont {Klautau}}, \bibinfo {author} {\bibfnamefont
  {O.}~\bibnamefont {Eriksson}},\ and\ \bibinfo {author} {\bibfnamefont
  {L.}~\bibnamefont {Nordstr{\"o}m}},\ }\bibfield  {title} {\bibinfo {title}
  {First-principles dzyaloshinskii--moriya interaction in a non-collinear
  framework},\ }\href@noop {} {\bibfield  {journal} {\bibinfo  {journal}
  {Scientific Reports}\ }\textbf {\bibinfo {volume} {10}},\ \bibinfo {pages}
  {20339} (\bibinfo {year} {2020})}\BibitemShut {NoStop}%
\bibitem [{\citenamefont {Ham}\ \emph {et~al.}(2021)\citenamefont {Ham},
  \citenamefont {Pradipto}, \citenamefont {Yakushiji}, \citenamefont {Kim},
  \citenamefont {Rhim}, \citenamefont {Nakamura}, \citenamefont {Shiota},
  \citenamefont {Kim},\ and\ \citenamefont {Ono}}]{ham2021dzyaloshinskii}%
  \BibitemOpen
  \bibfield  {author} {\bibinfo {author} {\bibfnamefont {W.~S.}\ \bibnamefont
  {Ham}}, \bibinfo {author} {\bibfnamefont {A.-M.}\ \bibnamefont {Pradipto}},
  \bibinfo {author} {\bibfnamefont {K.}~\bibnamefont {Yakushiji}}, \bibinfo
  {author} {\bibfnamefont {K.}~\bibnamefont {Kim}}, \bibinfo {author}
  {\bibfnamefont {S.~H.}\ \bibnamefont {Rhim}}, \bibinfo {author}
  {\bibfnamefont {K.}~\bibnamefont {Nakamura}}, \bibinfo {author}
  {\bibfnamefont {Y.}~\bibnamefont {Shiota}}, \bibinfo {author} {\bibfnamefont
  {S.}~\bibnamefont {Kim}},\ and\ \bibinfo {author} {\bibfnamefont
  {T.}~\bibnamefont {Ono}},\ }\bibfield  {title} {\bibinfo {title}
  {Dzyaloshinskii--moriya interaction in noncentrosymmetric superlattices},\
  }\href@noop {} {\bibfield  {journal} {\bibinfo  {journal} {npj Computational
  Materials}\ }\textbf {\bibinfo {volume} {7}},\ \bibinfo {pages} {129}
  (\bibinfo {year} {2021})}\BibitemShut {NoStop}%
\bibitem [{\citenamefont {Mahfouzi}\ and\ \citenamefont
  {Kioussis}(2021)}]{mahfouzi2021first}%
  \BibitemOpen
  \bibfield  {author} {\bibinfo {author} {\bibfnamefont {F.}~\bibnamefont
  {Mahfouzi}}\ and\ \bibinfo {author} {\bibfnamefont {N.}~\bibnamefont
  {Kioussis}},\ }\bibfield  {title} {\bibinfo {title} {First-principles
  calculation of the dzyaloshinskii-moriya interaction: A green's function
  approach},\ }\href@noop {} {\bibfield  {journal} {\bibinfo  {journal}
  {Physical Review B}\ }\textbf {\bibinfo {volume} {103}},\ \bibinfo {pages}
  {094410} (\bibinfo {year} {2021})}\BibitemShut {NoStop}%
\bibitem [{\citenamefont {Morshed}\ \emph {et~al.}(2021)\citenamefont
  {Morshed}, \citenamefont {Khoo}, \citenamefont {Quessab}, \citenamefont {Xu},
  \citenamefont {Laskowski}, \citenamefont {Balachandran}, \citenamefont
  {Kent},\ and\ \citenamefont {Ghosh}}]{morshed2021tuning}%
  \BibitemOpen
  \bibfield  {author} {\bibinfo {author} {\bibfnamefont {M.~G.}\ \bibnamefont
  {Morshed}}, \bibinfo {author} {\bibfnamefont {K.~H.}\ \bibnamefont {Khoo}},
  \bibinfo {author} {\bibfnamefont {Y.}~\bibnamefont {Quessab}}, \bibinfo
  {author} {\bibfnamefont {J.-W.}\ \bibnamefont {Xu}}, \bibinfo {author}
  {\bibfnamefont {R.}~\bibnamefont {Laskowski}}, \bibinfo {author}
  {\bibfnamefont {P.~V.}\ \bibnamefont {Balachandran}}, \bibinfo {author}
  {\bibfnamefont {A.~D.}\ \bibnamefont {Kent}},\ and\ \bibinfo {author}
  {\bibfnamefont {A.~W.}\ \bibnamefont {Ghosh}},\ }\bibfield  {title} {\bibinfo
  {title} {Tuning dzyaloshinskii-moriya interaction in ferrimagnetic gdco: A
  first-principles approach},\ }\href@noop {} {\bibfield  {journal} {\bibinfo
  {journal} {Physical Review B}\ }\textbf {\bibinfo {volume} {103}},\ \bibinfo
  {pages} {174414} (\bibinfo {year} {2021})}\BibitemShut {NoStop}%
\bibitem [{\citenamefont {Solovyev}(2023)}]{solovyev2023linear}%
  \BibitemOpen
  \bibfield  {author} {\bibinfo {author} {\bibfnamefont {I.}~\bibnamefont
  {Solovyev}},\ }\bibfield  {title} {\bibinfo {title} {Linear response based
  theories for dzyaloshinskii-moriya interactions},\ }\href@noop {} {\bibfield
  {journal} {\bibinfo  {journal} {Physical Review B}\ }\textbf {\bibinfo
  {volume} {107}},\ \bibinfo {pages} {054442} (\bibinfo {year}
  {2023})}\BibitemShut {NoStop}%
\bibitem [{\citenamefont {Allwood}\ \emph {et~al.}(2005)\citenamefont
  {Allwood}, \citenamefont {Xiong}, \citenamefont {Faulkner}, \citenamefont
  {Atkinson}, \citenamefont {Petit},\ and\ \citenamefont
  {Cowburn}}]{allwood2005magnetic}%
  \BibitemOpen
  \bibfield  {author} {\bibinfo {author} {\bibfnamefont {D.~A.}\ \bibnamefont
  {Allwood}}, \bibinfo {author} {\bibfnamefont {G.}~\bibnamefont {Xiong}},
  \bibinfo {author} {\bibfnamefont {C.}~\bibnamefont {Faulkner}}, \bibinfo
  {author} {\bibfnamefont {D.}~\bibnamefont {Atkinson}}, \bibinfo {author}
  {\bibfnamefont {D.}~\bibnamefont {Petit}},\ and\ \bibinfo {author}
  {\bibfnamefont {R.}~\bibnamefont {Cowburn}},\ }\bibfield  {title} {\bibinfo
  {title} {Magnetic domain-wall logic},\ }\href@noop {} {\bibfield  {journal}
  {\bibinfo  {journal} {science}\ }\textbf {\bibinfo {volume} {309}},\ \bibinfo
  {pages} {1688} (\bibinfo {year} {2005})}\BibitemShut {NoStop}%
\bibitem [{\citenamefont {Imre}\ \emph {et~al.}(2006)\citenamefont {Imre},
  \citenamefont {Csaba}, \citenamefont {Ji}, \citenamefont {Orlov},
  \citenamefont {Bernstein},\ and\ \citenamefont {Porod}}]{imre2006majority}%
  \BibitemOpen
  \bibfield  {author} {\bibinfo {author} {\bibfnamefont {A.}~\bibnamefont
  {Imre}}, \bibinfo {author} {\bibfnamefont {G.}~\bibnamefont {Csaba}},
  \bibinfo {author} {\bibfnamefont {L.}~\bibnamefont {Ji}}, \bibinfo {author}
  {\bibfnamefont {A.}~\bibnamefont {Orlov}}, \bibinfo {author} {\bibfnamefont
  {G.}~\bibnamefont {Bernstein}},\ and\ \bibinfo {author} {\bibfnamefont
  {W.}~\bibnamefont {Porod}},\ }\bibfield  {title} {\bibinfo {title} {Majority
  logic gate for magnetic quantum-dot cellular automata},\ }\href@noop {}
  {\bibfield  {journal} {\bibinfo  {journal} {Science}\ }\textbf {\bibinfo
  {volume} {311}},\ \bibinfo {pages} {205} (\bibinfo {year}
  {2006})}\BibitemShut {NoStop}%
\bibitem [{\citenamefont {Hrabec}\ \emph {et~al.}(2020)\citenamefont {Hrabec},
  \citenamefont {Luo}, \citenamefont {Heyderman},\ and\ \citenamefont
  {Gambardella}}]{hrabec2020synthetic}%
  \BibitemOpen
  \bibfield  {author} {\bibinfo {author} {\bibfnamefont {A.}~\bibnamefont
  {Hrabec}}, \bibinfo {author} {\bibfnamefont {Z.}~\bibnamefont {Luo}},
  \bibinfo {author} {\bibfnamefont {L.~J.}\ \bibnamefont {Heyderman}},\ and\
  \bibinfo {author} {\bibfnamefont {P.}~\bibnamefont {Gambardella}},\
  }\bibfield  {title} {\bibinfo {title} {Synthetic chiral magnets promoted by
  the dzyaloshinskii--moriya interaction},\ }\href@noop {} {\bibfield
  {journal} {\bibinfo  {journal} {Applied Physics Letters}\ }\textbf {\bibinfo
  {volume} {117}} (\bibinfo {year} {2020})}\BibitemShut {NoStop}%
\bibitem [{\citenamefont {Zhao}\ \emph {et~al.}(2021)\citenamefont {Zhao},
  \citenamefont {Chen}, \citenamefont {Prosandeev}, \citenamefont {Artyukhin},\
  and\ \citenamefont {Bellaiche}}]{zhao2021dzyaloshinskii}%
  \BibitemOpen
  \bibfield  {author} {\bibinfo {author} {\bibfnamefont {H.~J.}\ \bibnamefont
  {Zhao}}, \bibinfo {author} {\bibfnamefont {P.}~\bibnamefont {Chen}}, \bibinfo
  {author} {\bibfnamefont {S.}~\bibnamefont {Prosandeev}}, \bibinfo {author}
  {\bibfnamefont {S.}~\bibnamefont {Artyukhin}},\ and\ \bibinfo {author}
  {\bibfnamefont {L.}~\bibnamefont {Bellaiche}},\ }\bibfield  {title} {\bibinfo
  {title} {Dzyaloshinskii--moriya-like interaction in ferroelectrics and
  antiferroelectrics},\ }\href@noop {} {\bibfield  {journal} {\bibinfo
  {journal} {Nature Materials}\ }\textbf {\bibinfo {volume} {20}},\ \bibinfo
  {pages} {341} (\bibinfo {year} {2021})}\BibitemShut {NoStop}%
\bibitem [{\citenamefont {Gusev}\ \emph {et~al.}(2020)\citenamefont {Gusev},
  \citenamefont {Sadovnikov}, \citenamefont {Nikitov}, \citenamefont
  {Sapozhnikov},\ and\ \citenamefont {Udalov}}]{gusev2020manipulation}%
  \BibitemOpen
  \bibfield  {author} {\bibinfo {author} {\bibfnamefont {N.}~\bibnamefont
  {Gusev}}, \bibinfo {author} {\bibfnamefont {A.}~\bibnamefont {Sadovnikov}},
  \bibinfo {author} {\bibfnamefont {S.}~\bibnamefont {Nikitov}}, \bibinfo
  {author} {\bibfnamefont {M.}~\bibnamefont {Sapozhnikov}},\ and\ \bibinfo
  {author} {\bibfnamefont {O.}~\bibnamefont {Udalov}},\ }\bibfield  {title}
  {\bibinfo {title} {Manipulation of the dzyaloshinskii--moriya interaction in
  co/pt multilayers with strain},\ }\href@noop {} {\bibfield  {journal}
  {\bibinfo  {journal} {Physical review letters}\ }\textbf {\bibinfo {volume}
  {124}},\ \bibinfo {pages} {157202} (\bibinfo {year} {2020})}\BibitemShut
  {NoStop}%
\bibitem [{\citenamefont {Akanda}\ \emph {et~al.}(2020)\citenamefont {Akanda},
  \citenamefont {Park},\ and\ \citenamefont {Lake}}]{akanda2020interfacial}%
  \BibitemOpen
  \bibfield  {author} {\bibinfo {author} {\bibfnamefont {M.~R.~K.}\
  \bibnamefont {Akanda}}, \bibinfo {author} {\bibfnamefont {I.~J.}\
  \bibnamefont {Park}},\ and\ \bibinfo {author} {\bibfnamefont {R.~K.}\
  \bibnamefont {Lake}},\ }\bibfield  {title} {\bibinfo {title} {Interfacial
  dzyaloshinskii-moriya interaction of antiferromagnetic materials},\
  }\href@noop {} {\bibfield  {journal} {\bibinfo  {journal} {Physical Review
  B}\ }\textbf {\bibinfo {volume} {102}},\ \bibinfo {pages} {224414} (\bibinfo
  {year} {2020})}\BibitemShut {NoStop}%
\bibitem [{\citenamefont {Park}\ \emph {et~al.}(2020)\citenamefont {Park},
  \citenamefont {Kim}, \citenamefont {Nam}, \citenamefont {Jeon}, \citenamefont
  {Park}, \citenamefont {Kim}, \citenamefont {Lee}, \citenamefont {Min},\ and\
  \citenamefont {Choe}}]{park2020interfacial}%
  \BibitemOpen
  \bibfield  {author} {\bibinfo {author} {\bibfnamefont {Y.-K.}\ \bibnamefont
  {Park}}, \bibinfo {author} {\bibfnamefont {J.-S.}\ \bibnamefont {Kim}},
  \bibinfo {author} {\bibfnamefont {Y.-S.}\ \bibnamefont {Nam}}, \bibinfo
  {author} {\bibfnamefont {S.}~\bibnamefont {Jeon}}, \bibinfo {author}
  {\bibfnamefont {J.-H.}\ \bibnamefont {Park}}, \bibinfo {author}
  {\bibfnamefont {K.-W.}\ \bibnamefont {Kim}}, \bibinfo {author} {\bibfnamefont
  {H.-W.}\ \bibnamefont {Lee}}, \bibinfo {author} {\bibfnamefont {B.-C.}\
  \bibnamefont {Min}},\ and\ \bibinfo {author} {\bibfnamefont {S.-B.}\
  \bibnamefont {Choe}},\ }\bibfield  {title} {\bibinfo {title} {Interfacial
  atomic layers for full emergence of interfacial dzyaloshinskii--moriya
  interaction},\ }\href@noop {} {\bibfield  {journal} {\bibinfo  {journal} {NPG
  Asia Materials}\ }\textbf {\bibinfo {volume} {12}},\ \bibinfo {pages} {38}
  (\bibinfo {year} {2020})}\BibitemShut {NoStop}%
\bibitem [{\citenamefont {Legrand}\ \emph {et~al.}(2022)\citenamefont
  {Legrand}, \citenamefont {Sassi}, \citenamefont {Ajejas}, \citenamefont
  {Collin}, \citenamefont {Bocher}, \citenamefont {Jia}, \citenamefont
  {Hoffmann}, \citenamefont {Zimmermann}, \citenamefont {Bl{\"u}gel},
  \citenamefont {Reyren} \emph {et~al.}}]{legrand2022spatial}%
  \BibitemOpen
  \bibfield  {author} {\bibinfo {author} {\bibfnamefont {W.}~\bibnamefont
  {Legrand}}, \bibinfo {author} {\bibfnamefont {Y.}~\bibnamefont {Sassi}},
  \bibinfo {author} {\bibfnamefont {F.}~\bibnamefont {Ajejas}}, \bibinfo
  {author} {\bibfnamefont {S.}~\bibnamefont {Collin}}, \bibinfo {author}
  {\bibfnamefont {L.}~\bibnamefont {Bocher}}, \bibinfo {author} {\bibfnamefont
  {H.}~\bibnamefont {Jia}}, \bibinfo {author} {\bibfnamefont {M.}~\bibnamefont
  {Hoffmann}}, \bibinfo {author} {\bibfnamefont {B.}~\bibnamefont
  {Zimmermann}}, \bibinfo {author} {\bibfnamefont {S.}~\bibnamefont
  {Bl{\"u}gel}}, \bibinfo {author} {\bibfnamefont {N.}~\bibnamefont {Reyren}},
  \emph {et~al.},\ }\bibfield  {title} {\bibinfo {title} {Spatial extent of the
  dzyaloshinskii-moriya interaction at metallic interfaces},\ }\href@noop {}
  {\bibfield  {journal} {\bibinfo  {journal} {Physical Review Materials}\
  }\textbf {\bibinfo {volume} {6}},\ \bibinfo {pages} {024408} (\bibinfo {year}
  {2022})}\BibitemShut {NoStop}%
\bibitem [{\citenamefont {Shi}\ \emph {et~al.}(2017)\citenamefont {Shi},
  \citenamefont {Yuan}, \citenamefont {Mao}, \citenamefont {Ma},\ and\
  \citenamefont {Zhao}}]{shi2017robust}%
  \BibitemOpen
  \bibfield  {author} {\bibinfo {author} {\bibfnamefont {X.}~\bibnamefont
  {Shi}}, \bibinfo {author} {\bibfnamefont {H.}~\bibnamefont {Yuan}}, \bibinfo
  {author} {\bibfnamefont {X.}~\bibnamefont {Mao}}, \bibinfo {author}
  {\bibfnamefont {Y.}~\bibnamefont {Ma}},\ and\ \bibinfo {author}
  {\bibfnamefont {H.}~\bibnamefont {Zhao}},\ }\bibfield  {title} {\bibinfo
  {title} {Robust quantum state transfer inspired by dzyaloshinskii-moriya
  interactions},\ }\href@noop {} {\bibfield  {journal} {\bibinfo  {journal}
  {Physical Review A}\ }\textbf {\bibinfo {volume} {95}},\ \bibinfo {pages}
  {052332} (\bibinfo {year} {2017})}\BibitemShut {NoStop}%
\bibitem [{\citenamefont {Son}\ \emph {et~al.}(2019)\citenamefont {Son},
  \citenamefont {Park}, \citenamefont {Kim}, \citenamefont {Cho}, \citenamefont
  {Kim}, \citenamefont {Sandilands}, \citenamefont {Sohn}, \citenamefont
  {Park}, \citenamefont {Moon},\ and\ \citenamefont
  {Noh}}]{son2019unconventional}%
  \BibitemOpen
  \bibfield  {author} {\bibinfo {author} {\bibfnamefont {J.}~\bibnamefont
  {Son}}, \bibinfo {author} {\bibfnamefont {B.~C.}\ \bibnamefont {Park}},
  \bibinfo {author} {\bibfnamefont {C.~H.}\ \bibnamefont {Kim}}, \bibinfo
  {author} {\bibfnamefont {H.}~\bibnamefont {Cho}}, \bibinfo {author}
  {\bibfnamefont {S.~Y.}\ \bibnamefont {Kim}}, \bibinfo {author} {\bibfnamefont
  {L.~J.}\ \bibnamefont {Sandilands}}, \bibinfo {author} {\bibfnamefont
  {C.}~\bibnamefont {Sohn}}, \bibinfo {author} {\bibfnamefont {J.-G.}\
  \bibnamefont {Park}}, \bibinfo {author} {\bibfnamefont {S.~J.}\ \bibnamefont
  {Moon}},\ and\ \bibinfo {author} {\bibfnamefont {T.~W.}\ \bibnamefont
  {Noh}},\ }\bibfield  {title} {\bibinfo {title} {Unconventional spin-phonon
  coupling via the dzyaloshinskii--moriya interaction},\ }\href@noop {}
  {\bibfield  {journal} {\bibinfo  {journal} {npj Quantum materials}\ }\textbf
  {\bibinfo {volume} {4}},\ \bibinfo {pages} {17} (\bibinfo {year}
  {2019})}\BibitemShut {NoStop}%
\bibitem [{\citenamefont {Ozaydin}\ and\ \citenamefont
  {Altintas}(2020)}]{ozaydin2020parameter}%
  \BibitemOpen
  \bibfield  {author} {\bibinfo {author} {\bibfnamefont {F.}~\bibnamefont
  {Ozaydin}}\ and\ \bibinfo {author} {\bibfnamefont {A.~A.}\ \bibnamefont
  {Altintas}},\ }\bibfield  {title} {\bibinfo {title} {Parameter estimation
  with dzyaloshinskii--moriya interaction under external magnetic fields},\
  }\href@noop {} {\bibfield  {journal} {\bibinfo  {journal} {Optical and
  Quantum Electronics}\ }\textbf {\bibinfo {volume} {52}},\ \bibinfo {pages}
  {70} (\bibinfo {year} {2020})}\BibitemShut {NoStop}%
\bibitem [{\citenamefont {Khlifi}\ \emph {et~al.}(2020)\citenamefont {Khlifi},
  \citenamefont {El~Allati}, \citenamefont {Salah},\ and\ \citenamefont
  {Hassouni}}]{khlifi2020quantum}%
  \BibitemOpen
  \bibfield  {author} {\bibinfo {author} {\bibfnamefont {Y.}~\bibnamefont
  {Khlifi}}, \bibinfo {author} {\bibfnamefont {A.}~\bibnamefont {El~Allati}},
  \bibinfo {author} {\bibfnamefont {A.}~\bibnamefont {Salah}},\ and\ \bibinfo
  {author} {\bibfnamefont {Y.}~\bibnamefont {Hassouni}},\ }\bibfield  {title}
  {\bibinfo {title} {Quantum heat engine based on spin isotropic heisenberg
  models with dzyaloshinskii--moriya interaction},\ }\href@noop {} {\bibfield
  {journal} {\bibinfo  {journal} {International Journal of Modern Physics B}\
  }\textbf {\bibinfo {volume} {34}},\ \bibinfo {pages} {2050212} (\bibinfo
  {year} {2020})}\BibitemShut {NoStop}%
\bibitem [{\citenamefont {Hou{\c{c}}a}\ \emph {et~al.}(2022)\citenamefont
  {Hou{\c{c}}a}, \citenamefont {Belouad}, \citenamefont {Choubabi},
  \citenamefont {Kamal},\ and\ \citenamefont
  {El~Bouziani}}]{houcca2022quantum}%
  \BibitemOpen
  \bibfield  {author} {\bibinfo {author} {\bibfnamefont {R.}~\bibnamefont
  {Hou{\c{c}}a}}, \bibinfo {author} {\bibfnamefont {A.}~\bibnamefont
  {Belouad}}, \bibinfo {author} {\bibfnamefont {E.~B.}\ \bibnamefont
  {Choubabi}}, \bibinfo {author} {\bibfnamefont {A.}~\bibnamefont {Kamal}},\
  and\ \bibinfo {author} {\bibfnamefont {M.}~\bibnamefont {El~Bouziani}},\
  }\bibfield  {title} {\bibinfo {title} {Quantum teleportation via a two-qubit
  heisenberg xxx chain with x-component of dzyaloshinskii--moriya
  interaction},\ }\href@noop {} {\bibfield  {journal} {\bibinfo  {journal}
  {Journal of Magnetism and Magnetic Materials}\ }\textbf {\bibinfo {volume}
  {563}},\ \bibinfo {pages} {169816} (\bibinfo {year} {2022})}\BibitemShut
  {NoStop}%
\bibitem [{\citenamefont {Motamedifar}\ \emph {et~al.}(2023)\citenamefont
  {Motamedifar}, \citenamefont {Sadeghi},\ and\ \citenamefont
  {Golshani}}]{motamedifar2023entanglement}%
  \BibitemOpen
  \bibfield  {author} {\bibinfo {author} {\bibfnamefont {M.}~\bibnamefont
  {Motamedifar}}, \bibinfo {author} {\bibfnamefont {F.}~\bibnamefont
  {Sadeghi}},\ and\ \bibinfo {author} {\bibfnamefont {M.}~\bibnamefont
  {Golshani}},\ }\bibfield  {title} {\bibinfo {title} {Entanglement
  transmission due to the dzyaloshinskii--moriya interaction},\ }\href@noop {}
  {\bibfield  {journal} {\bibinfo  {journal} {Scientific Reports}\ }\textbf
  {\bibinfo {volume} {13}},\ \bibinfo {pages} {2932} (\bibinfo {year}
  {2023})}\BibitemShut {NoStop}%
\bibitem [{\citenamefont {Zhu}\ \emph {et~al.}(2023)\citenamefont {Zhu},
  \citenamefont {Wang}, \citenamefont {Shao}, \citenamefont {Zou},\ and\
  \citenamefont {Wu}}]{zhu2023effect}%
  \BibitemOpen
  \bibfield  {author} {\bibinfo {author} {\bibfnamefont {Z.-R.}\ \bibnamefont
  {Zhu}}, \bibinfo {author} {\bibfnamefont {Q.}~\bibnamefont {Wang}}, \bibinfo
  {author} {\bibfnamefont {B.}~\bibnamefont {Shao}}, \bibinfo {author}
  {\bibfnamefont {J.}~\bibnamefont {Zou}},\ and\ \bibinfo {author}
  {\bibfnamefont {L.-A.}\ \bibnamefont {Wu}},\ }\bibfield  {title} {\bibinfo
  {title} {Effect of the dzyaloshinskii-moriya interaction on quantum speed
  limit and orthogonality catastrophe},\ }\href@noop {} {\bibfield  {journal}
  {\bibinfo  {journal} {Physical Review A}\ }\textbf {\bibinfo {volume}
  {107}},\ \bibinfo {pages} {042427} (\bibinfo {year} {2023})}\BibitemShut
  {NoStop}%
\bibitem [{\citenamefont {Radhakrishnan}\ \emph {et~al.}(2017)\citenamefont
  {Radhakrishnan}, \citenamefont {Ermakov},\ and\ \citenamefont
  {Byrnes}}]{radhakrishnan2017quantum}%
  \BibitemOpen
  \bibfield  {author} {\bibinfo {author} {\bibfnamefont {C.}~\bibnamefont
  {Radhakrishnan}}, \bibinfo {author} {\bibfnamefont {I.}~\bibnamefont
  {Ermakov}},\ and\ \bibinfo {author} {\bibfnamefont {T.}~\bibnamefont
  {Byrnes}},\ }\bibfield  {title} {\bibinfo {title} {Quantum coherence of
  planar spin models with dzyaloshinsky-moriya interaction},\ }\href@noop {}
  {\bibfield  {journal} {\bibinfo  {journal} {Physical Review A}\ }\textbf
  {\bibinfo {volume} {96}},\ \bibinfo {pages} {012341} (\bibinfo {year}
  {2017})}\BibitemShut {NoStop}%
\bibitem [{\citenamefont {Liu}\ \emph {et~al.}(2011)\citenamefont {Liu},
  \citenamefont {Shao}, \citenamefont {Li}, \citenamefont {Zou},\ and\
  \citenamefont {Wu}}]{liu2011quantum}%
  \BibitemOpen
  \bibfield  {author} {\bibinfo {author} {\bibfnamefont {B.-Q.}\ \bibnamefont
  {Liu}}, \bibinfo {author} {\bibfnamefont {B.}~\bibnamefont {Shao}}, \bibinfo
  {author} {\bibfnamefont {J.-G.}\ \bibnamefont {Li}}, \bibinfo {author}
  {\bibfnamefont {J.}~\bibnamefont {Zou}},\ and\ \bibinfo {author}
  {\bibfnamefont {L.-A.}\ \bibnamefont {Wu}},\ }\bibfield  {title} {\bibinfo
  {title} {Quantum and classical correlations in the one-dimensional xy model
  with dzyaloshinskii-moriya interaction},\ }\href@noop {} {\bibfield
  {journal} {\bibinfo  {journal} {Physical Review A}\ }\textbf {\bibinfo
  {volume} {83}},\ \bibinfo {pages} {052112} (\bibinfo {year}
  {2011})}\BibitemShut {NoStop}%
\bibitem [{\citenamefont {Liu}\ \emph {et~al.}(2017)\citenamefont {Liu},
  \citenamefont {Wang}, \citenamefont {Sun},\ and\ \citenamefont
  {Ye}}]{liu2017quantum}%
  \BibitemOpen
  \bibfield  {author} {\bibinfo {author} {\bibfnamefont {C.-c.}\ \bibnamefont
  {Liu}}, \bibinfo {author} {\bibfnamefont {D.}~\bibnamefont {Wang}}, \bibinfo
  {author} {\bibfnamefont {W.-y.}\ \bibnamefont {Sun}},\ and\ \bibinfo {author}
  {\bibfnamefont {L.}~\bibnamefont {Ye}},\ }\bibfield  {title} {\bibinfo
  {title} {Quantum \uppercase {F}isher information, quantum entanglement and
  correlation close to quantum critical phenomena},\ }\href@noop {} {\bibfield
  {journal} {\bibinfo  {journal} {Quantum Information Processing}\ }\textbf
  {\bibinfo {volume} {16}},\ \bibinfo {pages} {1} (\bibinfo {year}
  {2017})}\BibitemShut {NoStop}%
\bibitem [{\citenamefont {Brida}\ \emph {et~al.}(2010)\citenamefont {Brida},
  \citenamefont {Degiovanni}, \citenamefont {Florio}, \citenamefont {Genovese},
  \citenamefont {Giorda}, \citenamefont {Meda}, \citenamefont {Paris},\ and\
  \citenamefont {Shurupov}}]{giorda1}%
  \BibitemOpen
  \bibfield  {author} {\bibinfo {author} {\bibfnamefont {G.}~\bibnamefont
  {Brida}}, \bibinfo {author} {\bibfnamefont {I.~P.}\ \bibnamefont
  {Degiovanni}}, \bibinfo {author} {\bibfnamefont {A.}~\bibnamefont {Florio}},
  \bibinfo {author} {\bibfnamefont {M.}~\bibnamefont {Genovese}}, \bibinfo
  {author} {\bibfnamefont {P.}~\bibnamefont {Giorda}}, \bibinfo {author}
  {\bibfnamefont {A.}~\bibnamefont {Meda}}, \bibinfo {author} {\bibfnamefont
  {M.~G.~A.}\ \bibnamefont {Paris}},\ and\ \bibinfo {author} {\bibfnamefont
  {A.}~\bibnamefont {Shurupov}},\ }\bibfield  {title} {\bibinfo {title}
  {Experimental estimation of entanglement at the quantum limit},\ }\href
  {https://doi.org/10.1103/PhysRevLett.104.100501} {\bibfield  {journal}
  {\bibinfo  {journal} {Phys. Rev. Lett.}\ }\textbf {\bibinfo {volume} {104}},\
  \bibinfo {pages} {100501} (\bibinfo {year} {2010})}\BibitemShut {NoStop}%
\bibitem [{\citenamefont {Genoni}\ \emph {et~al.}(2008)\citenamefont {Genoni},
  \citenamefont {Giorda},\ and\ \citenamefont {Paris}}]{giorda2}%
  \BibitemOpen
  \bibfield  {author} {\bibinfo {author} {\bibfnamefont {M.~G.}\ \bibnamefont
  {Genoni}}, \bibinfo {author} {\bibfnamefont {P.}~\bibnamefont {Giorda}},\
  and\ \bibinfo {author} {\bibfnamefont {M.~G.~A.}\ \bibnamefont {Paris}},\
  }\bibfield  {title} {\bibinfo {title} {Optimal estimation of entanglement},\
  }\href {https://doi.org/10.1103/PhysRevA.78.032303} {\bibfield  {journal}
  {\bibinfo  {journal} {Phys. Rev. A}\ }\textbf {\bibinfo {volume} {78}},\
  \bibinfo {pages} {032303} (\bibinfo {year} {2008})}\BibitemShut {NoStop}%
\bibitem [{\citenamefont {Teklu}\ \emph {et~al.}(2009)\citenamefont {Teklu},
  \citenamefont {Olivares},\ and\ \citenamefont {Paris}}]{teklu3}%
  \BibitemOpen
  \bibfield  {author} {\bibinfo {author} {\bibfnamefont {B.}~\bibnamefont
  {Teklu}}, \bibinfo {author} {\bibfnamefont {S.}~\bibnamefont {Olivares}},\
  and\ \bibinfo {author} {\bibfnamefont {M.~G.~A.}\ \bibnamefont {Paris}},\
  }\bibfield  {title} {\bibinfo {title} {Bayesian estimation of one-parameter
  qubit gates},\ }\href {https://doi.org/10.1088/0953-4075/42/3/035502}
  {\bibfield  {journal} {\bibinfo  {journal} {Journal of Physics B: Atomic,
  Molecular and Optical Physics}\ }\textbf {\bibinfo {volume} {42}},\ \bibinfo
  {pages} {035502} (\bibinfo {year} {2009})}\BibitemShut {NoStop}%
\bibitem [{\citenamefont {Rossi}\ and\ \citenamefont
  {Paris}(2015)}]{PhysRevA.92.010302}%
  \BibitemOpen
  \bibfield  {author} {\bibinfo {author} {\bibfnamefont {M.~A.~C.}\
  \bibnamefont {Rossi}}\ and\ \bibinfo {author} {\bibfnamefont {M.~G.~A.}\
  \bibnamefont {Paris}},\ }\bibfield  {title} {\bibinfo {title} {Entangled
  quantum probes for dynamical environmental noise},\ }\href
  {https://doi.org/10.1103/PhysRevA.92.010302} {\bibfield  {journal} {\bibinfo
  {journal} {Phys. Rev. A}\ }\textbf {\bibinfo {volume} {92}},\ \bibinfo
  {pages} {010302} (\bibinfo {year} {2015})}\BibitemShut {NoStop}%
\bibitem [{\citenamefont {Sommers}\ and\ \citenamefont
  {Zyczkowski}(2003)}]{sommers2003bures}%
  \BibitemOpen
  \bibfield  {author} {\bibinfo {author} {\bibfnamefont {H.-J.}\ \bibnamefont
  {Sommers}}\ and\ \bibinfo {author} {\bibfnamefont {K.}~\bibnamefont
  {Zyczkowski}},\ }\bibfield  {title} {\bibinfo {title} {Bures volume of the
  set of mixed quantum states},\ }\href@noop {} {\bibfield  {journal} {\bibinfo
   {journal} {Journal of Physics A: Mathematical and General}\ }\textbf
  {\bibinfo {volume} {36}},\ \bibinfo {pages} {10083} (\bibinfo {year}
  {2003})}\BibitemShut {NoStop}%
\bibitem [{\citenamefont {Paris}(2009)}]{paris2009quantum}%
  \BibitemOpen
  \bibfield  {author} {\bibinfo {author} {\bibfnamefont {M.~G.}\ \bibnamefont
  {Paris}},\ }\bibfield  {title} {\bibinfo {title} {Quantum estimation for
  quantum technology},\ }\href@noop {} {\bibfield  {journal} {\bibinfo
  {journal} {International Journal of Quantum Information}\ }\textbf {\bibinfo
  {volume} {7}},\ \bibinfo {pages} {125} (\bibinfo {year} {2009})}\BibitemShut
  {NoStop}%
\bibitem [{\citenamefont {Gu}(2010)}]{gu2010fidelity}%
  \BibitemOpen
  \bibfield  {author} {\bibinfo {author} {\bibfnamefont {S.-J.}\ \bibnamefont
  {Gu}},\ }\bibfield  {title} {\bibinfo {title} {Fidelity approach to quantum
  phase transitions},\ }\href@noop {} {\bibfield  {journal} {\bibinfo
  {journal} {International Journal of Modern Physics B}\ }\textbf {\bibinfo
  {volume} {24}},\ \bibinfo {pages} {4371} (\bibinfo {year}
  {2010})}\BibitemShut {NoStop}%
\bibitem [{\citenamefont {Damski}(2016)}]{damski2016fidelity}%
  \BibitemOpen
  \bibfield  {author} {\bibinfo {author} {\bibfnamefont {B.}~\bibnamefont
  {Damski}},\ }\bibfield  {title} {\bibinfo {title} {Fidelity approach to
  quantum phase transitions in quantum ising model},\ }in\ \href@noop {} {\emph
  {\bibinfo {booktitle} {Quantum Criticality in Condensed Matter: Phenomena,
  Materials and Ideas in Theory and Experiment}}}\ (\bibinfo  {publisher}
  {World Scientific},\ \bibinfo {year} {2016})\ pp.\ \bibinfo {pages}
  {159--182}\BibitemShut {NoStop}%
\bibitem [{\citenamefont
  {{\v{S}}afr{\'a}nek}(2017)}]{vsafranek2017discontinuities}%
  \BibitemOpen
  \bibfield  {author} {\bibinfo {author} {\bibfnamefont {D.}~\bibnamefont
  {{\v{S}}afr{\'a}nek}},\ }\bibfield  {title} {\bibinfo {title}
  {Discontinuities of the quantum \uppercase {F}isher information and the bures
  metric},\ }\href@noop {} {\bibfield  {journal} {\bibinfo  {journal} {Physical
  Review A}\ }\textbf {\bibinfo {volume} {95}},\ \bibinfo {pages} {052320}
  (\bibinfo {year} {2017})}\BibitemShut {NoStop}%
\bibitem [{\citenamefont {Invernizzi}\ \emph {et~al.}(2008)\citenamefont
  {Invernizzi}, \citenamefont {Korbman}, \citenamefont {Venuti},\ and\
  \citenamefont {Paris}}]{invernizzi2008optimal}%
  \BibitemOpen
  \bibfield  {author} {\bibinfo {author} {\bibfnamefont {C.}~\bibnamefont
  {Invernizzi}}, \bibinfo {author} {\bibfnamefont {M.}~\bibnamefont {Korbman}},
  \bibinfo {author} {\bibfnamefont {L.~C.}\ \bibnamefont {Venuti}},\ and\
  \bibinfo {author} {\bibfnamefont {M.~G.}\ \bibnamefont {Paris}},\ }\bibfield
  {title} {\bibinfo {title} {Optimal quantum estimation in spin systems at
  criticality},\ }\href@noop {} {\bibfield  {journal} {\bibinfo  {journal}
  {Physical Review A}\ }\textbf {\bibinfo {volume} {78}},\ \bibinfo {pages}
  {042106} (\bibinfo {year} {2008})}\BibitemShut {NoStop}%
\bibitem [{\citenamefont {Sun}\ \emph {et~al.}(2010)\citenamefont {Sun},
  \citenamefont {Ma}, \citenamefont {Lu},\ and\ \citenamefont
  {Wang}}]{sun2010fisher}%
  \BibitemOpen
  \bibfield  {author} {\bibinfo {author} {\bibfnamefont {Z.}~\bibnamefont
  {Sun}}, \bibinfo {author} {\bibfnamefont {J.}~\bibnamefont {Ma}}, \bibinfo
  {author} {\bibfnamefont {X.-M.}\ \bibnamefont {Lu}},\ and\ \bibinfo {author}
  {\bibfnamefont {X.}~\bibnamefont {Wang}},\ }\bibfield  {title} {\bibinfo
  {title} {Fisher information in a quantum-critical environment},\ }\href@noop
  {} {\bibfield  {journal} {\bibinfo  {journal} {Physical Review A}\ }\textbf
  {\bibinfo {volume} {82}},\ \bibinfo {pages} {022306} (\bibinfo {year}
  {2010})}\BibitemShut {NoStop}%
\bibitem [{\citenamefont {Carollo}\ \emph {et~al.}(2020)\citenamefont
  {Carollo}, \citenamefont {Valenti},\ and\ \citenamefont
  {Spagnolo}}]{carollo2020geometry}%
  \BibitemOpen
  \bibfield  {author} {\bibinfo {author} {\bibfnamefont {A.}~\bibnamefont
  {Carollo}}, \bibinfo {author} {\bibfnamefont {D.}~\bibnamefont {Valenti}},\
  and\ \bibinfo {author} {\bibfnamefont {B.}~\bibnamefont {Spagnolo}},\
  }\bibfield  {title} {\bibinfo {title} {Geometry of quantum phase
  transitions},\ }\href@noop {} {\bibfield  {journal} {\bibinfo  {journal}
  {Physics Reports}\ }\textbf {\bibinfo {volume} {838}},\ \bibinfo {pages} {1}
  (\bibinfo {year} {2020})}\BibitemShut {NoStop}%
\bibitem [{\citenamefont {Chu}\ \emph {et~al.}(2021)\citenamefont {Chu},
  \citenamefont {Zhang}, \citenamefont {Yu},\ and\ \citenamefont
  {Cai}}]{chu2021dynamic}%
  \BibitemOpen
  \bibfield  {author} {\bibinfo {author} {\bibfnamefont {Y.}~\bibnamefont
  {Chu}}, \bibinfo {author} {\bibfnamefont {S.}~\bibnamefont {Zhang}}, \bibinfo
  {author} {\bibfnamefont {B.}~\bibnamefont {Yu}},\ and\ \bibinfo {author}
  {\bibfnamefont {J.}~\bibnamefont {Cai}},\ }\bibfield  {title} {\bibinfo
  {title} {Dynamic framework for criticality-enhanced quantum sensing},\
  }\href@noop {} {\bibfield  {journal} {\bibinfo  {journal} {Physical Review
  Letters}\ }\textbf {\bibinfo {volume} {126}},\ \bibinfo {pages} {010502}
  (\bibinfo {year} {2021})}\BibitemShut {NoStop}%
\bibitem [{\citenamefont {Mihailescu}\ \emph {et~al.}(2023)\citenamefont
  {Mihailescu}, \citenamefont {Bayat}, \citenamefont {Campbell},\ and\
  \citenamefont {Mitchell}}]{mihailescu2023}%
  \BibitemOpen
  \bibfield  {author} {\bibinfo {author} {\bibfnamefont {G.}~\bibnamefont
  {Mihailescu}}, \bibinfo {author} {\bibfnamefont {A.}~\bibnamefont {Bayat}},
  \bibinfo {author} {\bibfnamefont {S.}~\bibnamefont {Campbell}},\ and\
  \bibinfo {author} {\bibfnamefont {A.~K.}\ \bibnamefont {Mitchell}},\
  }\href@noop {} {\bibinfo {title} {Multiparameter critical quantum metrology
  with impurity probes}} (\bibinfo {year} {2023}),\ \Eprint
  {https://arxiv.org/abs/2311.16931} {arXiv:2311.16931 [quant-ph]} \BibitemShut
  {NoStop}%
\bibitem [{\citenamefont {Maroufi}\ \emph {et~al.}(2021)\citenamefont
  {Maroufi}, \citenamefont {Laghmach}, \citenamefont {El~Hadfi},\ and\
  \citenamefont {Daoud}}]{maroufi2021analytical}%
  \BibitemOpen
  \bibfield  {author} {\bibinfo {author} {\bibfnamefont {B.}~\bibnamefont
  {Maroufi}}, \bibinfo {author} {\bibfnamefont {R.}~\bibnamefont {Laghmach}},
  \bibinfo {author} {\bibfnamefont {H.}~\bibnamefont {El~Hadfi}},\ and\
  \bibinfo {author} {\bibfnamefont {M.}~\bibnamefont {Daoud}},\ }\bibfield
  {title} {\bibinfo {title} {On the analytical derivation of quantum \uppercase
  {F}isher information and skew information for two qubit x states},\
  }\href@noop {} {\bibfield  {journal} {\bibinfo  {journal} {International
  Journal of Theoretical Physics}\ }\textbf {\bibinfo {volume} {60}},\ \bibinfo
  {pages} {3103} (\bibinfo {year} {2021})}\BibitemShut {NoStop}%
\bibitem [{\citenamefont {Albarelli}\ \emph {et~al.}(2020)\citenamefont
  {Albarelli}, \citenamefont {Barbieri}, \citenamefont {Genoni},\ and\
  \citenamefont {Gianani}}]{ALBARELLI2020126311}%
  \BibitemOpen
  \bibfield  {author} {\bibinfo {author} {\bibfnamefont {F.}~\bibnamefont
  {Albarelli}}, \bibinfo {author} {\bibfnamefont {M.}~\bibnamefont {Barbieri}},
  \bibinfo {author} {\bibfnamefont {M.}~\bibnamefont {Genoni}},\ and\ \bibinfo
  {author} {\bibfnamefont {I.}~\bibnamefont {Gianani}},\ }\bibfield  {title}
  {\bibinfo {title} {A perspective on multiparameter quantum metrology: From
  theoretical tools to applications in quantum imaging},\ }\href@noop {}
  {\bibfield  {journal} {\bibinfo  {journal} {Physics Letters A}\ }\textbf
  {\bibinfo {volume} {384}},\ \bibinfo {pages} {126311} (\bibinfo {year}
  {2020})}\BibitemShut {NoStop}%
\end{thebibliography}%

\end{document}